\documentclass[longauth]{aa}  

\usepackage{graphicx}
\usepackage{txfonts}
\usepackage[colorlinks=true, allcolors=blue]{hyperref}
\usepackage{placeins}
\usepackage[rightcaption]{sidecap}
\usepackage[english]{babel}
\usepackage{array}
\usepackage{capt-of}
\usepackage[normalem]{ulem}
\usepackage{lscape}
\usepackage{multirow}

\usepackage{bm}

\usepackage{xspace}
\newcommand{\kms}{km~s$^{-1}$\xspace}
\newcommand{\mum}{$\mu$m\xspace}

\begin{document}

   \title{JOYS+: mid-infrared detection of gas-phase SO$_2$ emission\\in a low-mass protostar}

  \subtitle{The case of NGC~1333~IRAS2A: hot core or accretion shock?}

   \author{M. L. van Gelder\inst{1}
          \and
          M. E. Ressler\inst{2}
          \and
          E. F. van Dishoeck\inst{1,3}
          \and
          P. Nazari\inst{1}
          \and
          B. Tabone\inst{4}
          \and
          J. H. Black\inst{5}
          \and
          {\L}. Tychoniec\inst{6}
          \and
          L. Francis\inst{1}
          \and
          M. Barsony\inst{7}
          \and
          H. Beuther\inst{8}
          \and
          A. Caratti o Garatti\inst{9}
          \and
          Y. Chen\inst{1}
          \and
          C. Gieser\inst{3}
          \and
          V. J. M. le Gouellec\inst{10,11}
          \and
          P. J. Kavanagh\inst{12}
          \and
          P. D. Klaassen\inst{13}
          \and
          B. W. P. Lew\inst{14}
          \and
          H. Linnartz\inst{15}
          \and
          L. Majumdar\inst{16,17}
          \and
          G. Perotti\inst{8}
          \and
          W. R. M. Rocha\inst{1,15}
          }

   \institute{Leiden Observatory, Leiden University, PO Box 9513, 2300RA Leiden, The Netherlands\\
              \email{vgelder@strw.leidenuniv.nl}
         \and
             Jet Propulsion Laboratory, California Institute of Technology, 4800 Oak Grove Drive, Pasadena, CA 91109, USA
         \and
             Max Planck Institut f\"ur Extraterrestrische Physik (MPE), Giessenbachstrasse 1, 85748 Garching, Germany
         \and
             Universit\'e Paris-Saclay, CNRS, Institut d'Astrophysique Spatiale, 91405 Orsay, France
         \and
             Chalmers University of Technology, Department of Space, Earth and Environment, Onsala Space Observatory, 439 92 Onsala, Sweden
         \and
             European Southern Observatory, Karl-Schwarzschild-Strasse 2, 85748 Garching bei M\"nchen, Germany
         \and
             SETI Institute 189 Bernardo Avenue, 2nd Floor, Mountain View, CA 94043, USA
         \and
             Max Planck Institute for Astronomy, K\"onigstuhl 17, 69117 Heidelberg, Germany
         \and
             INAF-Osservatorio Astronomico di Capodimonte, Salita Moiariello 16, 80131 Napoli, Italy
         \and
             NASA Ames Research Center, Space Science and Astrobiology Division M.S. 245-6 Moffett Field, CA 94035, USA
         \and 
             NASA Postdoctoral Program Fellow
         \and
             Department of Experimental Physics, Maynooth University, Maynooth, Co Kildare, Ireland
         \and
             UK Astronomy Technology Centre, Royal Observatory Edinburgh, Blackford Hill, Edinburgh EH9 3HJ, UK
         \and
             Bay Area Environmental Research Institute and NASA Ames Research Center, Moffett Field, CA 94035, USA
         \and
             Laboratory for Astrophysics, Leiden Observatory, Leiden University, P.O. Box 9513, 2300 RA Leiden, The Netherlands
         \and
             School of Earth and Planetary Sciences, National Institute of Science Education and Research, Jatni 752050, Odisha, India
         \and
             Homi Bhabha National Institute, Training School Complex, Anushaktinagar, Mumbai 400094, India
             }

   \date{Received XXX; accepted XXX}

 
  \abstract
   {The Mid-InfraRed Instrument (MIRI) on the {\it James Webb} Space Telescope (JWST) has sharpened our infrared eyes toward the star formation process. Due to its unprecedented spatial and spectral resolution and sensitivity in the mid-infrared, JWST/MIRI can see through highly extincted protostellar envelopes and probe the warm inner regions. An abundant molecule in these warm inner regions is SO$_2$, which is a common tracer of both outflow and accretion shocks as well as hot core chemistry.}
   {This paper presents the first mid-infrared detection of gaseous SO$_2$ emission in an embedded low-mass protostellar system rich in complex molecules and aims to determine the physical origin of the SO$_2$ emission.}
   {JWST/MIRI observations taken with the Medium Resolution Spectrometer (MRS) of the low-mass protostellar binary NGC~1333~IRAS2A are presented from the JWST Observations of Young protoStars (JOYS+) program, revealing emission from the SO$_2$ $\nu_3$ asymmetric stretching mode at 7.35~\mum. Using simple slab models and assuming local thermodynamic equilibrium (LTE), the rotational temperature and total number of SO$_2$ molecules are derived. The results are compared to those derived from high-angular resolution SO$_2$ data on the same scales ($\sim50-100$~au) obtained with the Atacama Large Millimeter/submillimeter Array (ALMA). }
   {
   The SO$_2$ emission from the $\nu_3$ band is predominantly located on $\sim50-100$~au scales around the MIR continuum peak of the main component of the binary, IRAS2A1.
   A rotational temperature of $92\pm8$~K is derived from the $\nu_3$ lines. This is in good agreement with the rotational temperature derived from pure rotational lines in the vibrational ground state (i.e., $\nu$=0) with ALMA ($104\pm5$~K) which are extended over similar scales. 
   However, the emission of the $\nu_3$ lines in the MIRI-MRS spectrum is not in LTE given that the total number of molecules predicted by a LTE model is found to be a factor $2\times10^4$ higher than what is derived for the $\nu=0$ state from the ALMA data.
   This difference can be explained by a vibrational temperature that is $\sim100$~K higher than the derived rotational temperature of the $\nu=0$ state: $T_{\rm vib}\sim200$~K vs $T_{\rm rot}=104\pm5$~K. 
   The brightness temperature derived from the continuum around the $\nu_3$ band ($\sim7.35$~\mum) of SO$_2$ is $\sim180$~K, which confirms that the $\nu_3=1$ level is not collisionally populated but rather infrared pumped by scattered radiation.
   This is also consistent with the non-detection of the $\nu_2$ bending mode at $18-20$~\mum.
   The similar rotational temperature derived from the MIRI-MRS and ALMA data implies that they are in fact tracing the same molecular gas.
   The inferred abundance of SO$_2$, using the LTE fit to the lines of the vibrational ground state in the ALMA data, is $1.0\pm0.3\times10^{-8}$ with respect to H$_2$, which is on the lower side compared to interstellar and cometary ices ($10^{-8}-10^{-7}$).
   }
   {
   Given the rotational temperature, the extent of the emission ($\sim100$~au in radius), and the narrow line widths in the ALMA data ($\sim3.5$~km~s$^{-1}$), the SO$_2$ in IRAS2A likely originates from ice sublimation in the central hot core around the protostar rather than from an accretion shock at the disk-envelope boundary. 
   Furthermore, this paper shows the importance of radiative pumping and of combining JWST observations with those from millimeter interferometers such as ALMA to probe the physics on disk scales and to infer molecular abundances. 
   }


   \maketitle
%

\section{Introduction}
The embedded protostellar phase of star formation is very rich in terms of chemistry. The {\it James Webb} Space Telescope (JWST) provides unique new opportunities for probing these deeply embedded protostellar sources \citep[][]{Yang2022,Harsono2023,vanDishoeck2023,Beuther2023,Gieser2023}. An interesting element to study is sulfur, as the total volatile sulfur budget in protostars appears to be depleted by more than two orders of magnitude with respect to the diffuse clouds \citep{Ruffle1999}.
The sulfur likely resides in unobservable refractory reservoirs, sulfur-allotropes, or FeS inclusions \citep{Woods2015,Kama2019}. Even among the volatile species, the main sulfur-reservoir remains poorly constrained \citep[e.g.,][]{Drozdovskaya2018,Navarro-Almaida2020,Kushwahaa2023}. 
It is therefore important to constrain the physical conditions in which the volatile sulfur-bearing species reside in order to understand their chemistry and constrain the main sulfur reservoirs.
JWST has proven that it can detect sulfur-bearing species, both in interstellar ices \citep[e.g., SO$_2$, OCS;][]{Yang2022,McClure2023,Rochasubm} and even in exoplanetary atmospheres \citep[e.g., SO$_2$;][]{Tsai2023}. Here, we present one of the first medium-resolution mid-infrared (MIR) spectra of a low-mass Class~0 protostellar system, NGC~1333~IRAS2A, containing the first detection of gaseous SO$_2$ in emission at MIR wavelengths. 

One of the most detected sulfur-bearing species toward low-mass protostellar systems in millimeter observations is SO$_2$ \citep[e.g.,][]{ArturdelaVillarmois2023}. It is a shock tracer which is often present in outflow and jet shocks \citep[e.g.,][]{Blake1987,Codella2014,Taquet2020,Tychoniec2021}, where it is either sputtered from icy dust grains or formed through gas-phase chemistry \citep[e.g.,][]{PineaudesForets1993}. SO$_2$ is also often observed on smaller scales in accretion shocks at the disk-envelope boundary \citep[e.g.,][]{Sakai2014,Oya2019,ArturdelaVillarmois2019,ArturdelaVillarmois2022,Garufi2022}, where it is either formed through gas-phase chemistry or thermally sublimated from the ices \citep{Miura2017,vanGelder2021}. Furthermore, it is also a good tracer of disk winds \citep{Tabone2017} or hot core regions where the bulk of the ices are sublimating \citep[e.g.,][]{Drozdovskaya2018,Codella2021}. Most of these studies are based on submillimeter observations with interferometers such as the Atacama Large Millimeter/submillimeter array (ALMA) or the NOrthern Extended Millimeter Array (NOEMA) which can trace the pure rotational transitions of the vibrational ground state of SO$_2$ but are not able detect ro-vibrational transitions.

These ro-vibrational lines are best traced at mid-infrared (MIR, i.e.,$\sim5-30$~\mum) wavelengths. SO$_2$ has three fundamental vibrational modes: the $\nu_1$ symmetrical stretching mode around $8.5-9$~\mum, the $\nu_2$ bending mode around $18-20$~\mum, and the $\nu_3$ asymmetrical stretching mode around $7.2-7.4$~\mum \citep[][]{Briggs1970}. At MIR wavelengths, gaseous SO$_2$ has thus far only been observed in absorption toward high-mass protostellar systems \citep{Keane2001,Dungee2018,Nickerson2023}. These studies have shown that the SO$_2$ in these high-mass systems resides at typical temperatures of $\sim100-300$~K, although also higher temperatures of up to $700$~K have been reported \citep{Keane2001}. The average abundances with respect to H$_2$ are $>10^{-7}$, which is consistent with the SO$_2$ abundance of interstellar ices \citep[][]{Boogert1997,Boogert2015,Zasowski2009,McClure2023,Rochasubm} and cometary ices \citep{Altwegg2019,Rubin2019}, suggesting that gaseous SO$_2$ may originate from ice sublimation in the inner hot cores of these high-mass protostellar systems. A MIR detection of gaseous SO$_2$, either in absorption or emission, toward low-mass protostellar systems is, to the best of our knowledge, still lacking. 

Most studies at submillimeter wavelengths assume local thermodynamic equilibrium (LTE) in deriving the physical conditions (i.e., column density, excitation temperature) of SO$_2$, which is a good approximation given that most pure rotational levels can be collisionally populated at typical inner envelope densities ($10^6-10^8$~cm$^{-3}$) and can therefore be characterized by a single excitation temperature. However, when studying ro-vibrational lines at MIR wavelengths it is important to be aware of non-LTE effects. The critical densities of these ro-vibrational transitions are typically $>10^{10}$~cm$^{-3}$ \citep[e.g., for HCN and CO$_2$;][]{Bruderer2015,Bosman2017}, meaning that vibrationally excited levels are only collisionally populated in the inner $\lesssim1$~au around low-mass protostars. Furthermore, molecules can be pumped into a vibrationally excited state by a strong infrared radiation field which boost the line fluxes of MIR transitions far above from what is expected from collisional excitation \citep[e.g.,][]{Boonman2003,Sonnentrucker2007}.
For SO$_2$ no collisional rate coefficients are available for its ro-vibrational transitions in the MIR, therefore not allowing for a full non-LTE analysis, but the critical densities of its MIR transitions likely have similar high critical densities as HCN and CO$_2$. However, the importance of non-LTE effects can still be constrained through the comparison of ro-vibrational transitions detected at MIR wavelengths to pure rotational transitions in the vibrational ground state measured at submillimeter wavelengths. 

In this paper we present JWST/Mid-InfraRed Instrument \citep[MIRI;][]{Rieke2015,Wright2015,Wright2023} observations from the JWST Observations of Young protoStars (JOYS+) program, providing the first MIR detection of SO$_2$ in emission toward the low-mass protostellar system NGC~1333~IRAS2A (hereafter IRAS2A), and compare it to the results of high-angular resolution ALMA data of the same region. IRAS2A is a binary Class~0 system with a separation between IRAS2A1 (VLA1) and IRAS2A2 (VLA2) of $\sim0.6''$  \citep[i.e.,$\sim180$~au;][]{Tobin2015_IRAS2A,Tobin2016,Jorgensen2022} located in the Perseus star-forming region at a distance of about 293~pc \citep{Ortiz-Leon2018}. It hosts one of the most famous hot corinos \citep[e.g.,][]{Jorgensen2005,Bottinelli2007,Maury2014,Taquet2015}, and drives two powerful almost perpendicular large-scale outflows \citep[e.g.,][]{Arce2010,Tobin2015_IRAS2A,Taquet2020}. 
Recently, \citet{Jorgensen2022} showed that the binary interaction results in a misalignment in the outflow and infalling streamers around IRAS2A1. 
Furthermore, IRAS2A was part of the {\it Spitzer} Cores to Disks (c2d) survey, revealing the main components of the ices \citep[][]{Boogert2008,Oberg2011}, but the Infrared Spectograph (IRS) of {\it Spitzer} only had a spectral resolving power of $R=60-120$ at the critical wavelength range of $5-10$~\mum and a large aperture ($>5''$). Here, we present MIRI observations taken with the Medium Resolution Spectrometer \citep[MRS;][]{Wells2015,Labiano2021,Argyriou2023,Jones2023} with a spectral resolving power of $R\sim3500$ at $5-10$~\mum and subarcsecond resolution ($\sim0.2-0.4''$). This is one of the first JWST/MIRI-MRS spectra of a low-mass Class~0 protostellar system \citep[see also e.g., IRAS~15398-3359;][]{Yang2022}.

This paper is structured as follows. In Sect.~\ref{sec:observations}, the JWST/MIRI and ALMA observations are described together with the method of fitting the SO$_2$ emission. The results are presented in Sect.~\ref{sec:results} and discussed in Sect.~\ref{sec:discussion}. Our main conclusions are summarized in Sect.~\ref{sec:conclusions}.


\section{Observations and analysis}
\label{sec:observations}
\subsection{Observations}
\subsubsection{MIRI-MRS}
The MIRI-MRS observations were carried out as part of the Cycle 1 Guaranteed Time Program (GTO) 1236 (PI: M. E. Ressler) on January 9th 2023 with a 2-point dither pattern optimized for extended sources. A dedicated background observation was also taken with 2-point dither pattern to allow for a proper subtraction of the telescope background and detector artifacts. In both cases, the FASTR1 readout mode was used with all three gratings (A, B, C) in all four Channels, providing the full wavelength coverage of MIRI-MRS (4.9--28.6~$\mu$m). The integration time in each grating was 111~s, resulting in a total integration time of 333~s. All observations included simultaneous off-source imaging using the F1500W filter.

\begin{figure*}
    \centering
    \includegraphics[width=\linewidth]{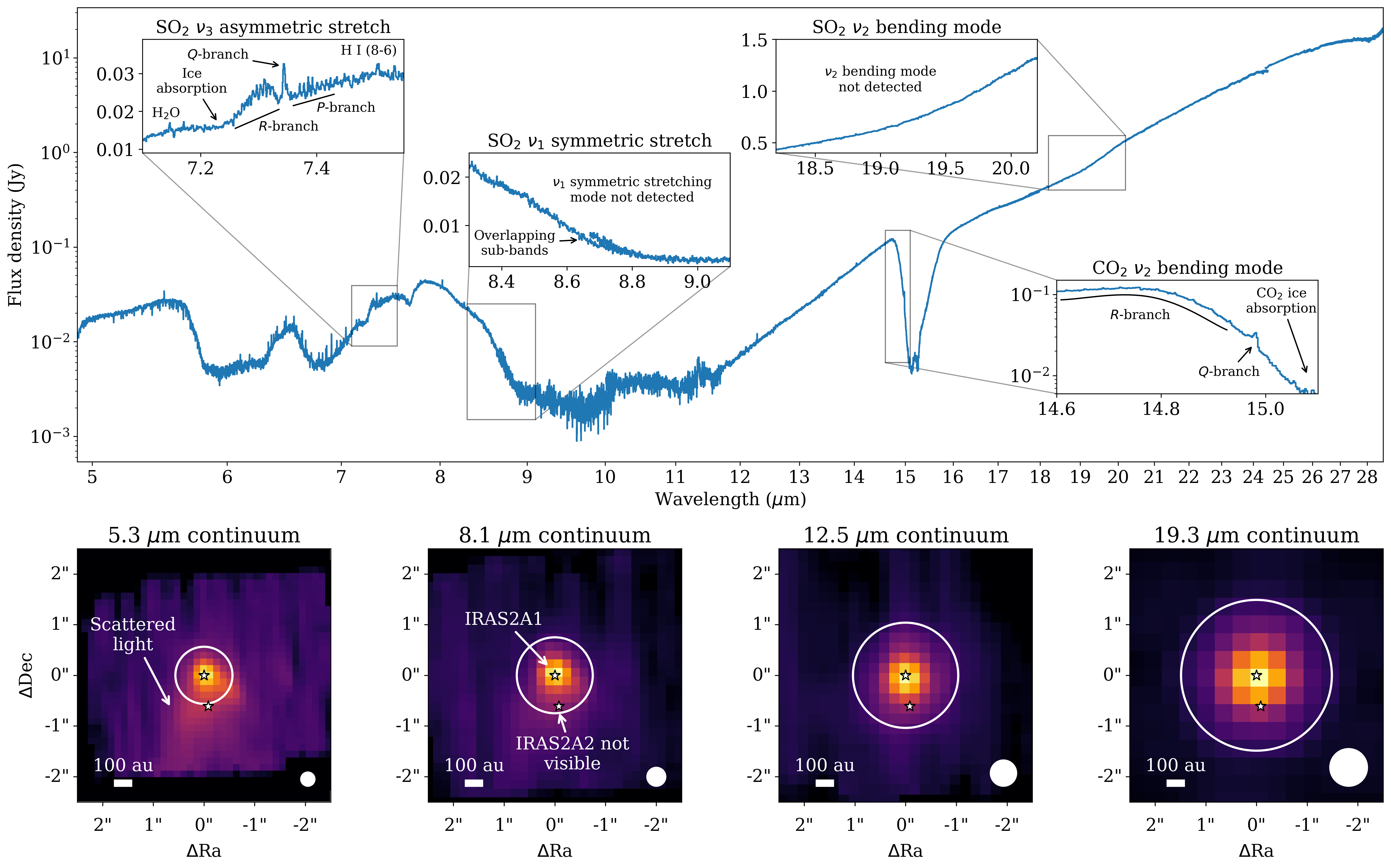}
    \caption{Spectrum (top panel) and continuum images at various wavelengths (bottom row) of IRAS2A. In the top panel, four insets are presented around the SO$_2$ $\nu_3$ asymmetric stretching mode ($\sim7.35$~\mum), SO$_2$ $\nu_1$ symmetric stretching mode ($\sim8.7$~\mum), SO$_2$ $\nu_2$ bending mode ($\sim19$~\mum), and CO$_2$ $\nu_2$ bending mode ($\sim15$~\mum), where emission from the latter originates mostly from the outflow (see Fig.~\ref{fig:co2_miri_map}). 
    The 12 sub-bands of the MIRI-MRS spectrum are not stitched and show minor offsets in overlapping wavelengths (e.g. inset on the SO$_2$ $\nu_1$ symmetric stretching mode).
    The bottom row shows from left to right the dust continuum around 5.3~\mum, 8.1~\mum, 12.5~\mum, and 19.3~\mum, using a {\tt sqrt} stretch to enhance fainter features without over-saturating bright emission. The open white circle indicates the size of the aperture from which the spectrum was extracted (i.e, $1.1''$, $1.5''$, $2.1''$, and $3.0''$ in diameter at 5.3~\mum, 8.1~\mum, 12.5~\mum, and 19.3~\mum, respectively), which increases as function of wavelength with the increase of the size of the PSF. A scale bar is displayed in the bottom left of each panel and the size of the PSF is presented as the filled white circle in the bottom right.}
    \label{fig:spec_and_cont_images}
\end{figure*}

The observations were processed through all three stages of the JWST calibration pipeline version 1.11.0 \citep{bushouse_howard_2023_8067394} using the reference context {\tt jwst$\_$1097.pmap} of the JWST Calibration Reference Data System \citep[CRDS;][]{Greenfield2016}. The raw data were first processed through the {\tt Detector1Pipeline} using the default settings. Following this, the {\tt Spec2Pipeline} was performed, including the correction for fringes with the fringe flat for extended sources (Mueller et al. in prep.) and applying the detector level residual fringe correction (Kavanagh et al. in prep). Furthermore, the background was also subtracted at this step using the rate files of the dedicated background. A bad-pixel routine was applied to the {\tt Spec2Pipeline} products outside of the default MIRI-MRS pipeline using the Vortex Image Processing package \citep[VIP;][]{Christiaens2023}. The data were further processed using the {\tt Spec3Pipeline} with both the master background and outlier rejection routines switched off. The background was already subtracted in the {\tt Spec2Pipeline} and the outlier rejection routine was skipped because it did not significantly improve the quality of the data. Moreover, the data cubes created with the {\tt Spec3Pipeline} were created for each band of each Channel separately.

The main component of the binary, IRAS2A1 \citep[VLA1;][]{Tobin2015_IRAS2A}, is clearly detected across all wavelengths whereas the companion protostar, IRAS2A2 (VLA2), is not detected (see Fig.~\ref{fig:spec_and_cont_images}). The reason for the non-detection could be because it is about an order of magnitude less massive and less luminous than IRAS2A1 \citep[][]{Tobin2016,Tobin2018}, or because it is more embedded. Therefore, only one spectrum was manually extracted from the peak of the mid-infrared continuum (R.A. (J200) 03$^{\rm h}$28$^{\rm m}$55.57$^{\rm s}$, Dec (J2000) 31$^{\rm d}$14$^{\rm m}$36.76$^{\rm s}$) at 5.5~$\mu$m (Channel 1A, see Fig.~\ref{fig:spec_and_cont_images}) and is assumed to only contain emission related to IRAS2A1 (hereafter IRAS2A). 
The diameter of the aperture was set to $4\times{\rm FWHM_{PSF}}$, with ${\rm FWHM_{PSF}}$ the empirically derived full width at half maximum of the MIRI-MRS point spread function \citep[${\rm FWHM_{PSF}} = 0.033 (\lambda/\mu{\rm m}) + 0.106 ''$, i.e., 0.35$''$ at 7.35~\mum;][]{Law2023}, in order to capture as much emission as possible without adding too much noise. An additional spectrum level residual fringe correction was applied primarily to remove the high-frequency dichroic fringes in Channels 3 and 4 (Kavanagh et al. in prep). No additional spectral stitching was applied between the bands.

\begin{figure*}
    \centering
    \includegraphics[width=\linewidth]{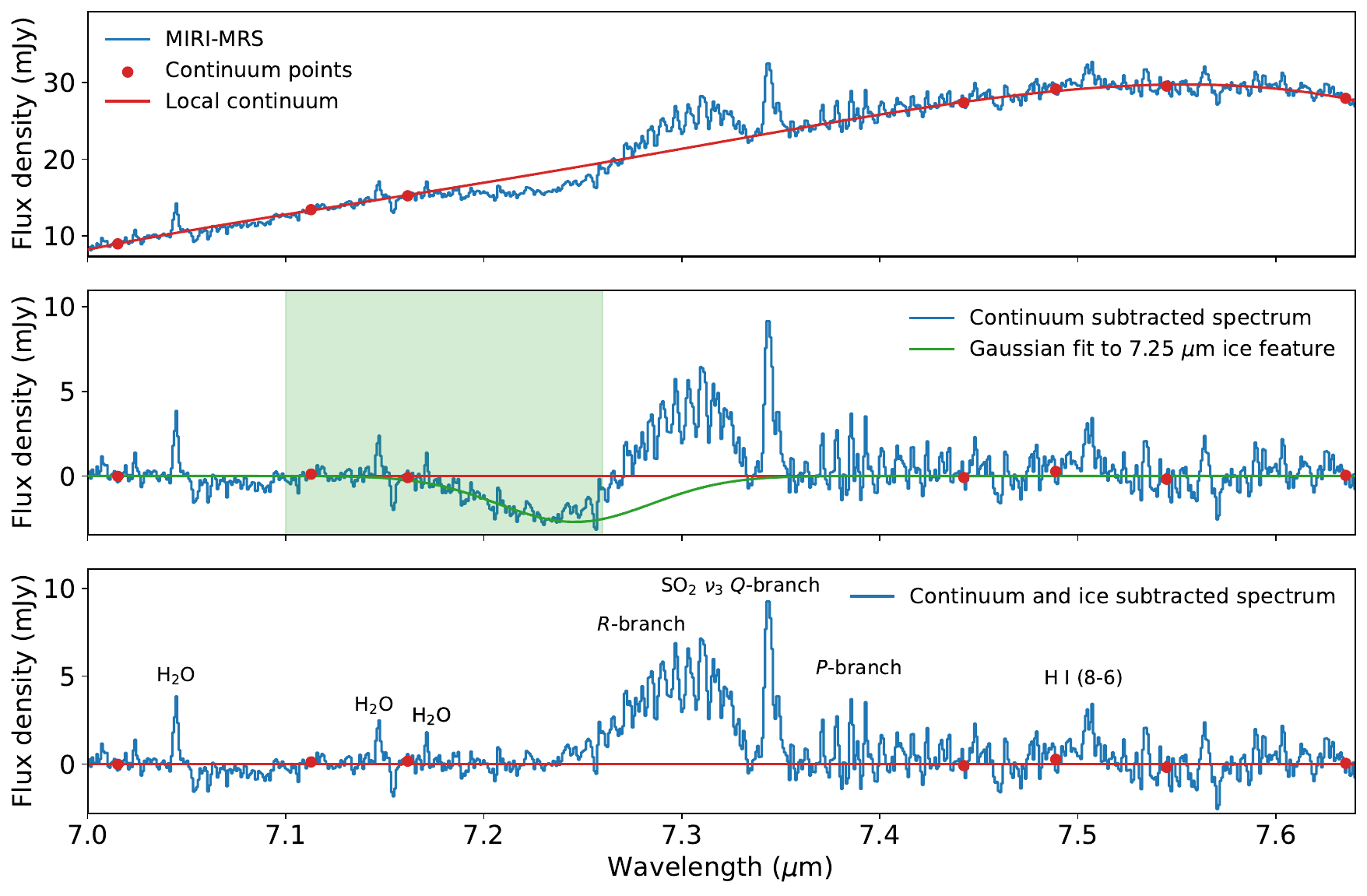}
    \caption{
    {\it Top}: the spectrum of IRAS2A around 7.35~\mum showing the gas-phase SO$_2$ emission superimposed at various ice absorption bands. The red line shows the estimated local continuum based on a fourth-order polynomial fit through the red dots.
    {\it Middle}: the spectrum of IRAS2A with the local continuum subtracted. The 7.25~\mum complex organics ice absorption feature is clearly present. A simple Gaussian fit to this ice absorption feature is shown in green based on fitting the spectrum in the shaded region.
    {\it Bottom}: The final continuum and ice feature subtracted spectrum. 
    }
    \label{fig:spec_contsubtract}
\end{figure*}

\subsubsection{ALMA data}
The ALMA data analyzed in this paper are taken from program 2021.1.01578.S (PI: B. Tabone), which targeted several Class~0 protostars in Perseus in Band~7, covering several transitions of SO$_2$ and $^{34}$SO$_2$. These data were taken in an extended configuration (C-6, $\theta_\mathrm{beam} = 0.12''\times0.09''$) and in a more compact configuration (C-3, $\theta_{\rm beam} = 0.58''\times0.34''$) to include larger-scale emission. 
All spectral windows have a velocity resolution of 0.22~km~s$^{-1}$ except for two windows which have a resolution of 0.44~km~s$^{-1}$ and one continuum window at 0.87~km~s$^{-1}$. The data were pipeline calibrated and imaged with the Common Astronomy Software Applications\footnote{\url{https://casa.nrao.edu/}} \citep[CASA;][]{McMullin2007} version 6.4.1.12. The continuum was subtracted with the \texttt{uvcontsub} task using carefully selected line-free channels. Following this, the two configurations were combined using the \texttt{tclean} task for a \texttt{briggs} weighting of 0.5 with a circular mask with a radius of $2''$ centered on the main continuum peak. The synthesized beam of the final data cubes is $\theta_\mathrm{beam} = 0.13''\times0.10''$.
In order for direct comparison to the MIRI data, spectra are extracted from a $1.4''$ diameter aperture centered on the continuum position of IRAS2A1. The noise level of the extracted spectrum is 0.015~Jy, and a flux calibration uncertainty of 5\% is assumed.

\subsection{Continuum subtraction around SO$_2$ $\nu_3$ band}
SO$_2$ is an asymmetric rotor and has three fundamental vibrational modes: a symmetrical stretching mode ($\nu_1=1151$~cm$^{-1}$, $\lambda\sim8.5-9.0$~\mum), a bending mode around ($\nu_2=518$~cm$^{-1}$, $\lambda\sim18-20$~\mum), and an asymmetrical stretching mode \citep[$\nu_3=1362$~cm$^{-1}$, $\lambda\sim7.2-7.4$~\mum;][]{Briggs1970,Person1982}. Around 7.35~\mum, clear molecular emission originating from the $\nu_3$ band is present (see inset in Fig.~\ref{fig:spec_and_cont_images}), but both the $\nu_1$ and $\nu_2$ bands are not detected. The analysis of the MIRI-MRS data will therefore be focused on the $\nu_3$ band and the absence of the $\nu_1$ and $\nu_2$ bands will be further discussed in Sect.~\ref{subsec:nu1_nu2}.

In order to fit the SO$_2$ emission in the $\nu_3$ band, the local continuum had to be subtracted. However, the spectral region surrounding the emission features is dominated by absorption of ices such as the 7.25~$\mu$m and 7.4~$\mu$m ice bands that are typically ascribed to complex organics including ethanol (CH$_3$CH$_2$OH) and acetaldehyde (CH$_3$CHO) or the formate ion \citep[HCOO$^-$; e.g.,][]{Schutte1999,Oberg2011,Boogert2015,TerwisschavanScheltinga2018,Rochasubm}. A fourth order polynomial was fitted through obvious line-free channels  (i.e., 7.015, 7.113, 7.162, 7.442, 7.489, 7.545, 7.635~\mum; see top panel of Fig.~\ref{fig:spec_contsubtract}) selected outside of the SO$_2$ emission range and the 7.25~$\mu$m and 7.4~$\mu$m ice bands to estimate the local continuum. Following the subtraction of this local continuum, a clear ice absorption feature remains around 7.25~$\mu$m. The generally equally strong 7.4~\mum ice feature appears to be absent, but this likely originates from the superposition with the $P$-branch of the SO$_2$ emission (see Sect.~\ref{sec:results}). To estimate the strength of the 7.25~$\mu$m ice absorption band, a simple Gaussian model was fitted to the part of the spectrum where no clear molecular emission features are present ($7.1-7.26$~\mum; green shaded area in Fig.~\ref{fig:spec_contsubtract}). Using this Gaussian fit, the contribution of the 7.25~$\mu$m ice feature was subtracted, providing a spectrum with only the contribution of the molecular emission (Fig.~\ref{fig:spec_contsubtract} bottom). 

\subsection{LTE slab model fitting}
\label{subsec:slab_model_fitting}
The molecular emission features in the continuum-subtracted spectra were fitted with simple slab models assuming local thermodynamic equilibrium (LTE). This is a similar approach to what is applied to other low- and high-mass sources \citep[][]{Carr2008,Salyk2011,Tabone2023,Grant2023,Perotti2023,Francissubm}. 
It is important to note here that the assumption of LTE is likely not valid for the MIR ro-vibrational transitions of SO$_2$ detected in IRAS2A. Nevertheless, LTE models can still constrain physical quantities such as the rotational temperature within a vibrationally excited state. The implications of the assumption of LTE will be further discussed in Sect.~\ref{subsec:non-LTE}.

In the slab models, the emission is assumed to arise from a slab of gas with an excitation temperature $T_{\rm ex}$ (which in the case of LTE is approximately equal to the kinetic temperature), a column density $N$, and an emitting area equal to $\pi R^2$. The latter is parameterized as a circular emitting area with a radius $R$, but could in reality originate from any shaped region with an area equal to $\pi R^2$. The intrinsic velocity broadening $\Delta V$ was fixed to 3.5~km~s$^{-1}$ based on ALMA data covering SO$_2$ (see Sect.~\ref{subsec:comp_to_ALMA}). It is important to note that in case of optically thin emission, the derived column densities are independent of $\Delta V$, whereas in case of optically thick emission the column density scales with 1/$\Delta V$ \citep{Tabone2023}. Given that SO$_2$ has many ro-vibrational lines close to each other in wavelength in the MIR, it is important to take line overlap into account in order to derive an accurate column density and excitation temperature. Moreover, the spectral resolving power of MIRI-MRS around 7.35~\mum is about $\lambda/\Delta\lambda\sim3500$ \citep{Labiano2021,Jones2023}, corresponding to a velocity resolution of $\sim85$~\kms, which means that the lines are not spectrally resolved and could be blended with other lines (of e.g., SO$_2$ itself, H$_2$O) in the MIRI data.

The spectroscopic information (i.e., line wavelengths, Einstein $A_{\rm ij}$, upper energy level $E_{\rm up}$, degeneracy $g_{\rm up}$) of the MIR ro-vibrational transitions of SO$_2$ was taken from the HITRAN database\footnote{\url{https://hitran.org/}} \citep{Gordon2022} and converted to the Leiden Atomic and Molecular Database (LAMDA) format \citep{vanderTak2020} in order to make it compatible with the slab model code. The partition function of SO$_2$ was calculated with the {\tt TIPS\_2021\_PYTHON} code provided by the HITRAN database. 

The best-fitting $N$ and $T_{\rm ex}$ were computed by creating a large grid covering $10^{14}-10^{19}$~cm$^{-2}$ in steps of 0.05 in $\log_{10}$ and $50-250$~K in steps of 1~K, respectively. Any higher column densities or temperatures were excluded by visual inspection of the spectrum. For each grid point, the LTE spectrum of SO$_2$ was calculated at a spectral resolving power of $R=3500$ and the best-fitting emitting area was computed by minimizing the $\chi^2$ \citep[see Appendix C of][for more details]{Grant2023}. Only selected channels (i.e.,$7.23-7.255$, $7.258-7.268$, $7.2755-7.3535$~\mum), that do not include obvious emission/absorption features related to other molecules (i.e., H$_2$O) or artifacts, are taken into account. Specifically, all wavelengths longer than $>7.3535$~\mum were excluded because of the contribution of the 7.4~\mum ice feature to the flux. Similarly, the noise level was estimated from line-free channels ($7.06-7.13$, $7.195-7.225$, $7.47-7.50$~\mum) to be about 0.59~mJy, and a flux calibration uncertainty of 5\% is assumed \citep{Argyriou2023}. The LTE models are corrected for an absolute extinction of $A_{\rm V} = 55$~mag \citep[estimated based on the depth of the silicate absorption feature;][see also Fig.~\ref{fig:cont_tau}]{Rochasubm} using the modified version of the extinction law of \citet{McClure2009} introduced in Appendix~\ref{app:ext_corr}. The effect of differential extinction caused by the 7.25~\mum and 7.4~\mum ice absorption on the molecular emission is assumed to be negligible. The best-fit parameters ($N$, $T_{\rm ex}$, and $R$) and their uncertainties were derived from the grid by minimizing the $\chi^2$.


\section{Results}
\label{sec:results}
\subsection{MIRI-MRS}
\label{subsec:res_MIRI}
The full spectrum of IRAS2A is presented in the top panel of Fig.~\ref{fig:spec_and_cont_images}. In the bottom panels of Fig.~\ref{fig:spec_and_cont_images}, the continuum maps are presented at four increasing wavelengths, each obtained at a different MIRI-MRS Channel. At the shortest wavelengths ($\sim5.3$~\mum; left panel), the continuum clearly shows scattered light from the blue-shifted outflow cavity extending toward the south. This scattered light is visible up to wavelengths of $\sim8$~\mum (middle left panel), but is no longer present at longer wavelengths ($>8.5$~\mum; two panels on the right). Only the main component of the binary, IRAS2A1 \citep[VLA1;][]{Tobin2015_IRAS2A}, is detected whereas the companion protostar, IRAS2A2 (VLA2), is not detected. The non-detection of IRAS2A could originate from it being an order of magnitude less massive and luminous \citep[][]{Tobin2016,Tobin2018}, or because it is more embedded.

\begin{figure}
    \centering
    \includegraphics[width=\linewidth]{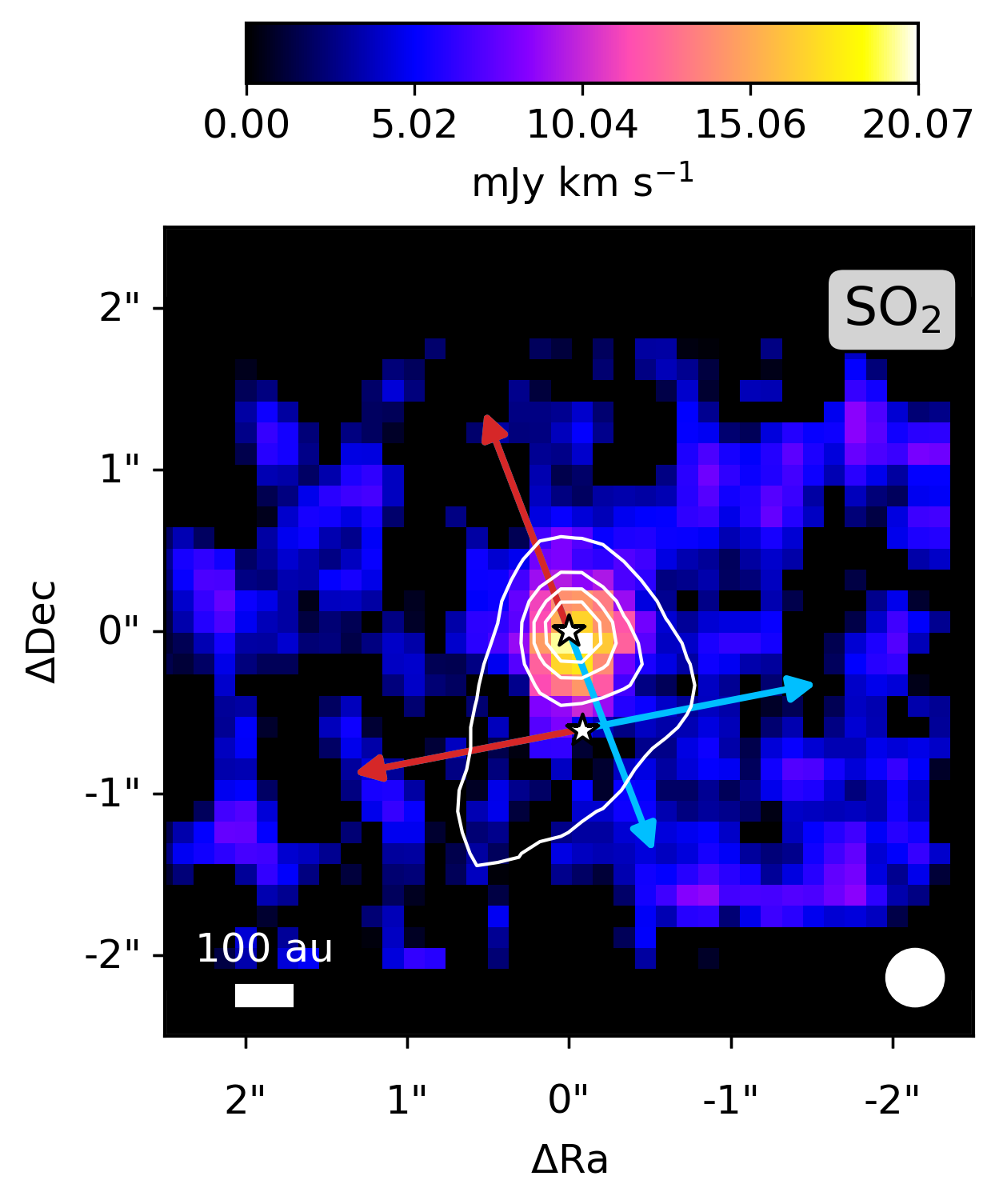}
    \caption{
    Integrated intensity map of SO$_2$ observed with MIRI-MRS in color. The image is integrated over the $Q$-branch of the $\nu_3$ band: [7.34,7.35]~\mum and shown using a {\tt sqrt} stretch to enhance fainter emission. The continuum around 7.35~\mum is overlaid in white contours and peaks on the primary component, IRAS2A1, with some extent toward the south from scattered light. The positions of IRAS2A1 and IRAS2A2 are depicted with the white stars. A white scale bar is displayed in the bottom left and the size of the PSF is presented as the filled white circle in the bottom right. The direction of the two outflows originating from IRAS2A1 and IRAS2A2 are indicated with the colored arrows \citep{Tobin2015_IRAS2A}. 
    }
    \label{fig:so2_miri_map}
\end{figure}

\begin{figure*}
    \centering
    \includegraphics[width=\linewidth]{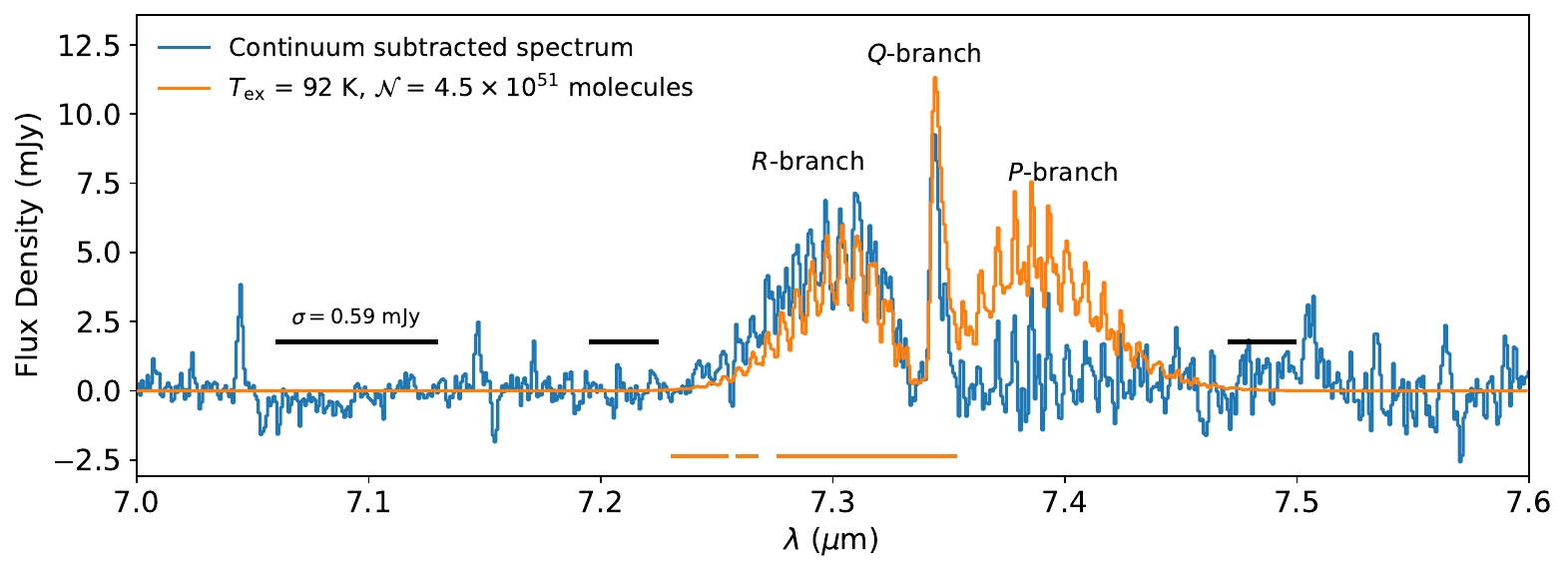}
    \caption{The continuum subtracted spectrum (blue) overlaid with the best-fit SO$_2$ model (orange). The orange bars in the bottom indicate the wavelength ranges that were included in the fit and the black bars show the wavelength ranges over which $\sigma$ is calculated. The $P$-branch lines around 7.4~\mum appear to be overfitted since the 7.4~\mum ice absorption could not be disentangled from the gas-phase emission. }
    \label{fig:so2_mod}
\end{figure*}

\begin{figure}
    \centering
    \includegraphics[width=\linewidth]{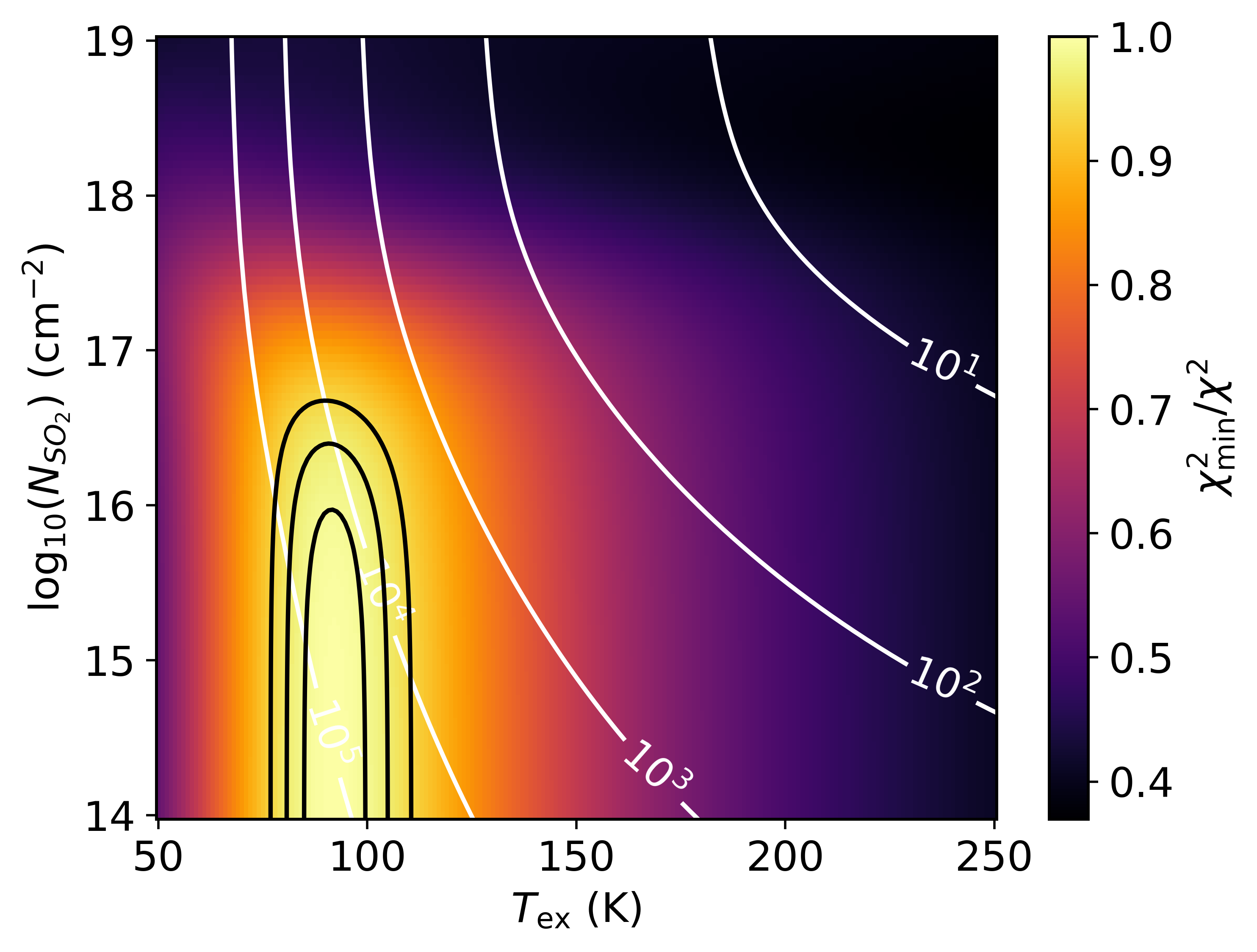}
    \caption{$\chi^2$ map of the LTE slab model fit of SO$_2$. The $\chi^2$ values are inversely normalized by the minimum $\chi^2$ and shown in color. 
    The 1, 2, and 3$\sigma$ confidence intervals are presented as the black contours. 
    The best-fitting emitting radius $R$ (in au) is represented by the white contours and is calculated for each grid point by minimizing the $\chi^2$ assuming a circular emitting area of $\pi R^2$. The $\chi^2$ map clearly reveals that the SO$_2$ emission is optically thin with an excitation temperature of $92 \pm 8$~K.} 
    \label{fig:chi2_map}
\end{figure}

The integrated intensity map over the SO$_2$ $\nu_3$ $Q$-branch in the (continuum subtracted) MIRI-MRS data is presented in Fig.~\ref{fig:so2_miri_map}. The SO$_2$ emission is mostly located around the continuum peak of IRAS2A1. After deconvolution with the FHWM of the PSF \citep[][]{Law2023}, the extent of SO$_2$ emission is $0.5''\times0.3''$ in diameter (i.e.,$\sim150\times90$~au). This very similar to the extent of H$_2$O emission lines that are likely tracing the inner hot corino (see Fig.~\ref{fig:co2_miri_map}). Some very weak extended SO$_2$ emission (on $\sim500$~au scales) is also present in the direction of the blue-shifted lobe of the north-south outflow \citep{Tobin2015_IRAS2A,Jorgensen2022}. 
However, the extended SO$_2$ emission could also be related to IRAS2A2, which is located about $\sim0.6''$ (i.e., $\sim180$~au) from IRAS2A1 to the south west \citep[][]{Tobin2015_IRAS2A,Tobin2016,Tobin2018,Jorgensen2022}.

The SO$_2$ emission is clearly tracing a more compact component than the CO$_2$  emission around 15~\mum, which peaks more toward the blue-shifted part of the north-south outflow than toward the central protostellar component (see Fig.~\ref{fig:co2_miri_map}). 
Similarly, the lower $S$-branch lines of H$_2$ (e.g., S(1)) are also peaking in the blue-shifted outflow, but higher $S$-branch lines (e.g., S(7)) originating from higher $E_{\rm up}$ levels are mostly peaking on the continuum position (see Fig.~\ref{fig:co2_miri_map}).
Furthermore, several ice absorption features including those of SO$_2$ ice are detected toward the bright continuum source which are analyzed by \citet{Rochasubm}. Any gas emission and absorption features other than SO$_2$ in this source, including the CO$_2$ emission in the outflow, will also be presented in future papers.

The best-fit LTE slab model to the $\nu_3$ band is presented in Fig.~\ref{fig:so2_mod} overlaid on the continuum subtracted spectrum. The fit slightly overestimates the \mbox{$Q$-branch} around 7.35~\mum but provides a very reasonable fit to the \mbox{$R$-branch} around 7.3~\mum. The \mbox{$P$-branch} lines around 7.4~\mum were not included in the fit because the contribution of the 7.4~\mum ice feature, which decreases the measured fluxes of the \mbox{$P$-branch} lines, could not be disentangled from the gas-phase emission of SO$_2$ a priori. Armed with a good model for the $Q$- and \mbox{$R$-branches}, Figure.~\ref{fig:so2_sub_spec} clearly shows that the local continuum was estimated too high in the 7.4~\mum region, proving that in fact the 7.4~\mum ice absorption feature is present. The emission of the SO$_2$ \mbox{$P$-branch} is almost equally strong as the absorption caused by the ices.

The $\chi^2$ map is shown in Fig.~\ref{fig:chi2_map}. Based on the $\chi^2$ map, the best-fitting excitation temperature can be accurately constrained (with a 1$\sigma$ error) at $T_{\rm ex} = 92 \pm 8$~K. Any lower excitation temperature results in a too strong and too narrow $Q$-branch compared to the $R$-branch whereas higher excitation temperatures result in a too broad $Q$-branch and also start overshooting the $R$-branch. 
Furthermore, the SO$_2$ emission is optically thin since optically thick emission would give a lower ratio between the measured $R$- and $Q$-branch fluxes.
The column density and emitting area are therefore completely degenerate with each other (i.e., the vertical profile of the contours in Fig.~\ref{fig:chi2_map}). Optically thick emission would appear as a more banana shaped profile in Fig.~\ref{fig:chi2_map} and allow for constraining both the column density and emitting area \citep[see e.g.,][]{Pontoppidan2002,Salyk2011,Grant2023,Tabone2023}. 
Given the degeneracy, no accurate column density of SO$_2$ can be derived, but the total number of molecules, $\mathcal{N} = N \pi R^2$, can be determined: $\mathcal{N} = 4.5 \pm 4.0 \times 10^{51}$~molecules. 
However, the data can only be accurately fitted for very large emitting area of $\gtrsim5000$~au, which is in strong contrast with the extent of the emission in Fig.~\ref{fig:so2_miri_map} ($<150$~au). 

The latter is a clear indication that the assumption of LTE is not valid and thus that derived physical quantities, most notably the number of molecules, should be analyzed in more detail. 
Since the rotational distribution within the $\nu_3$ band is well-fitted by a single temperature, this temperature accurately describes the rotational temperature within this vibrationally excited state, but it does not necessarily represent the kinetic temperature. On the other hand, the values derived for the total number of molecules and the size of the emitting area are not valid, both because of the conflict with the measured size of the emitting area as well as the large discrepancy with the values derived for the vibrational ground state in the ALMA data (see Sect.~\ref{subsec:comp_to_ALMA}). This will be further discussed in Section~\ref{subsec:non-LTE}.

\subsection{Comparison to ALMA data}
\label{subsec:comp_to_ALMA}
\begin{figure*}[t]
    \centering
    \includegraphics[width=0.33\linewidth]{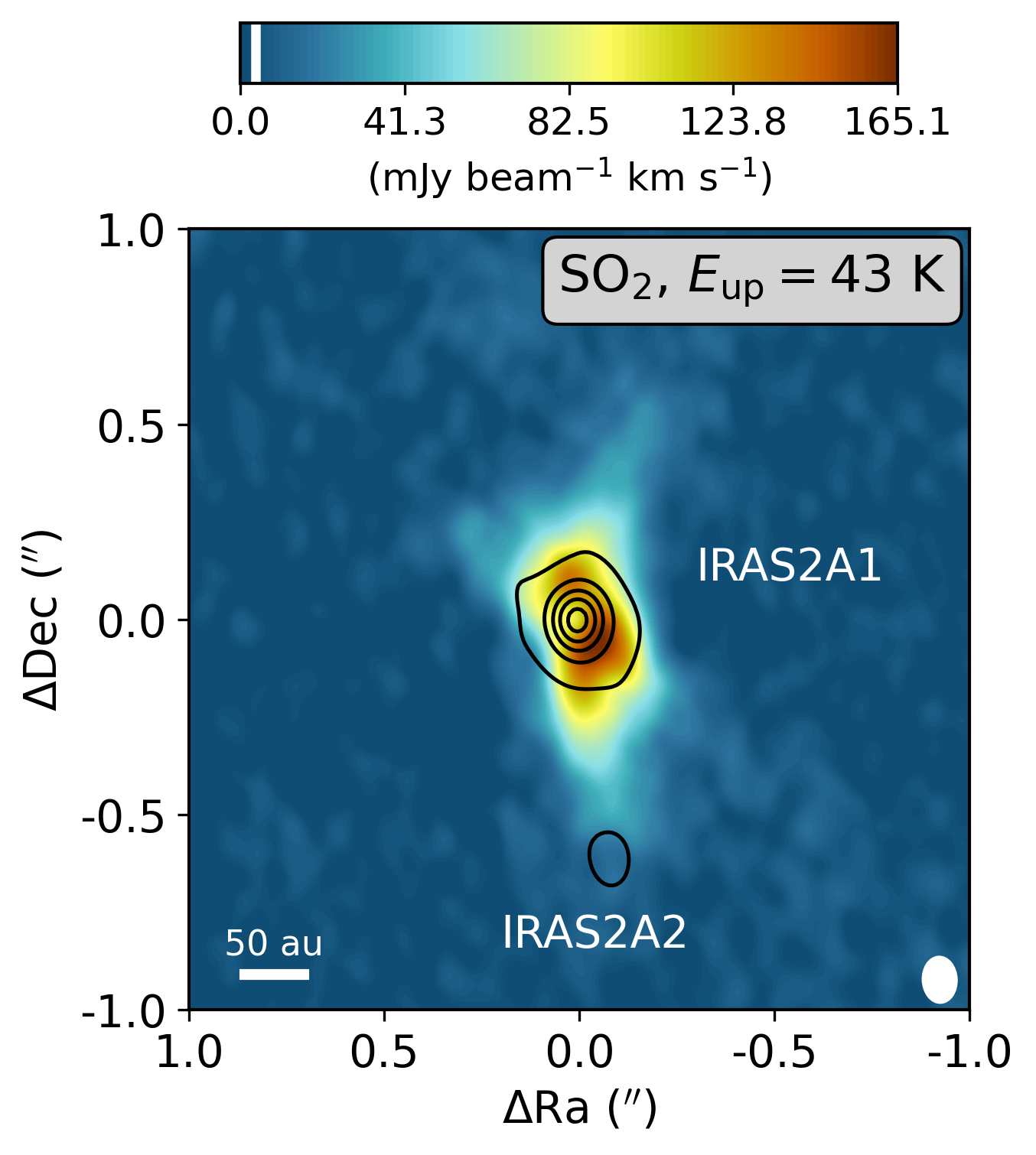}
    \includegraphics[width=0.33\linewidth]{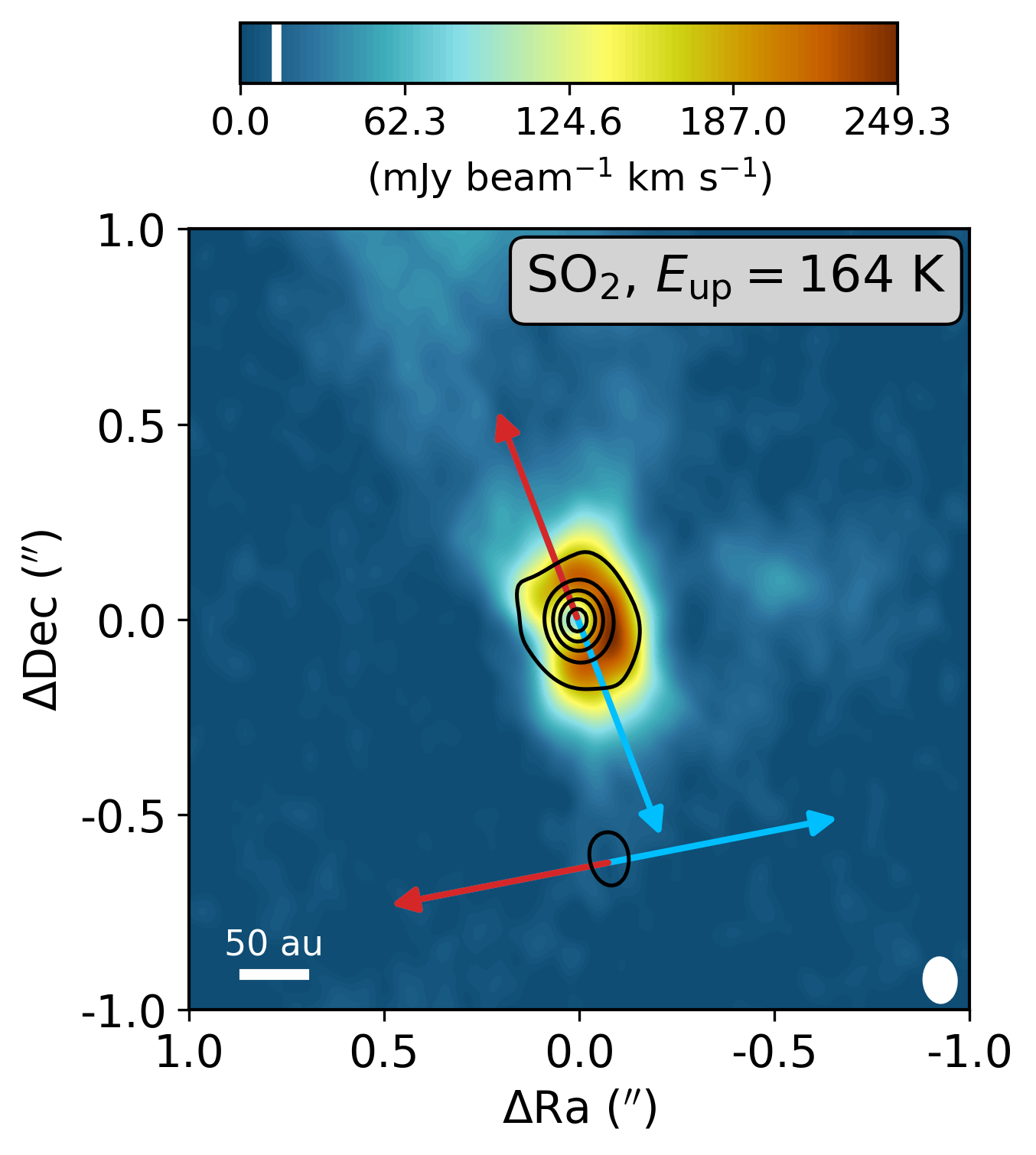}
    \includegraphics[width=0.33\linewidth]{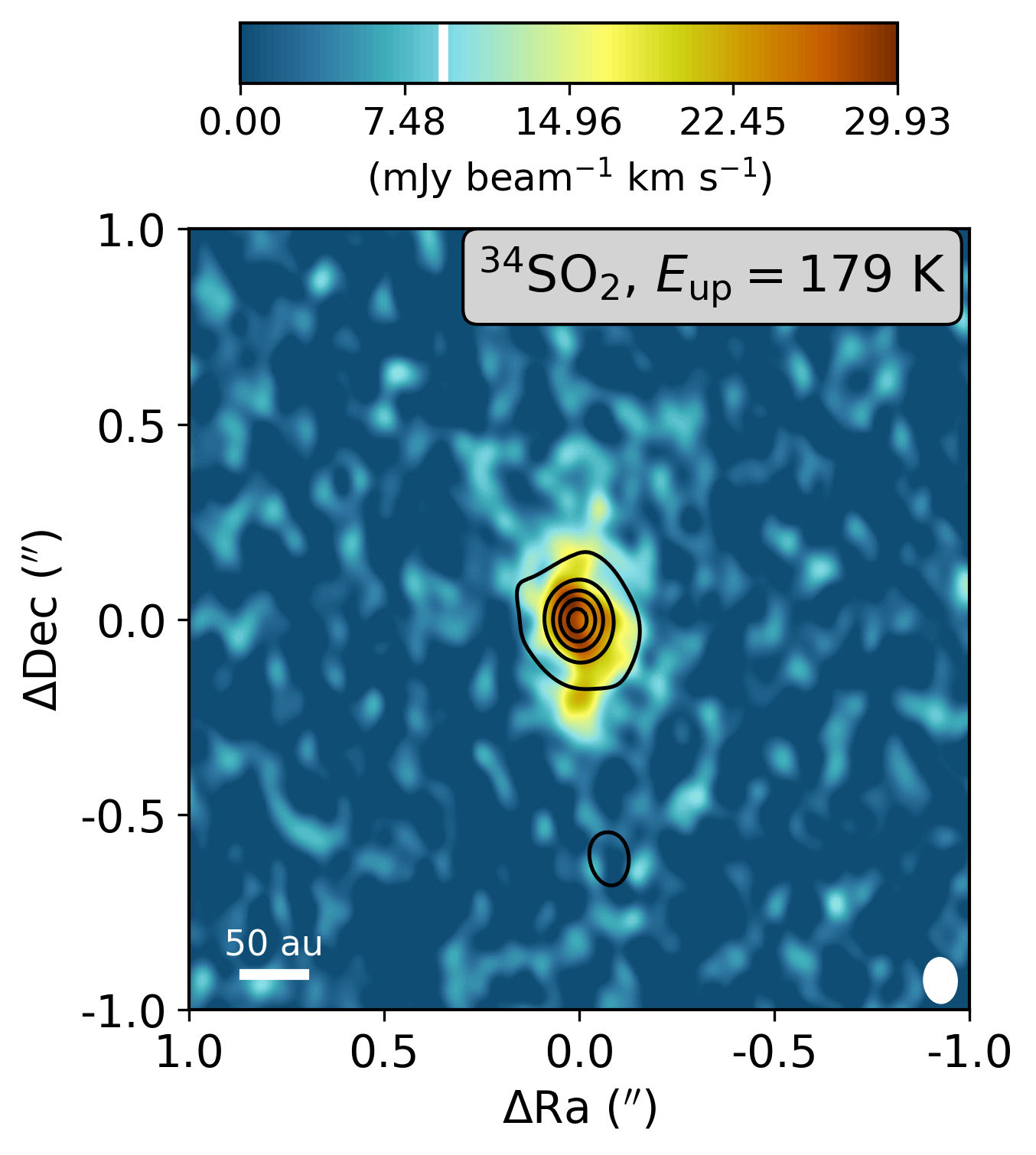}
    \caption{Integrated intensity maps of the SO$_2$ $8_{2,6}-7_{1,7}$ (left, $E_{\rm up} = 43$~K), SO$_2$ $16_{4,12}-16_{3,13}$ (middle, $E_{\rm up} = 164$~K), and $^{34}$SO$_2$ $17_{4,14}-17_{3,15}$ (right, $E_{\rm up} = 178$~K) transitions in color. 
    The images are integrated over [-2,2]~km~s$^{-1}$ with respect to the $V_{\rm lsr}$ of 6.7~km~s$^{-1}$. 
    The white vertical bar in the colorbar on top of each image indicates the 3$\sigma$ threshold. 
    The 0.875~mm continuum is overlaid in the black contours. The main continuum peak is associated with IRAS2A1 and the secondary peak toward the south with IRAS2A2. The direction of the two outflows originating from IRAS2A1 and IRAS2A2 are indicated with the colored arrows in the middle panel \citep{Tobin2015_IRAS2A}. The size of the beam is shown in the bottom right and in the bottom left a scale bar is displayed.} 
    \label{fig:ALMA_maps}
\end{figure*}

Integrated intensity maps of the SO$_2$ $8_{2,6}-7_{1,7}$ ($E_{\rm up} = 43$~K), SO$_2$ $16_{4,12}-16_{3,13}$ ($E_{\rm up} = 164$~K), and $^{34}$SO$_2$ $17_{4,14}-17_{3,15}$ ($E_{\rm up} = 178$~K) pure rotational transitions in their vibrational ground states (i.e., $\nu=0$) are presented in Fig.~\ref{fig:ALMA_maps}. The emission is spatially resolved and is extended in the north-west to south-east direction, following the direction of the outflow of IRAS2A1 \citep[e.g.,][]{Tobin2015_IRAS2A,Jorgensen2022}. However, most of the emission is concentrated in the inner $\sim1''$, similar to what is seen in the MIRI-MRS data (Fig.~\ref{fig:so2_miri_map}). After deconvolution with the beam, the size of the emission originating from the SO$_2$ $16_{4,12}-16_{3,13}$ transition is $\sim0.5''\times0.3''$ in diameter, corresponding to $\sim150\times90$~au. The SO$_2$ $8_{2,6}-7_{1,7}$ transition has a slightly larger deconvolved emitting size of $\sim0.6''\times0.4''$ ($\sim180\times120$~au) likely because its lower $E_{\rm up}$ makes it more sensitive to colder and more extended material in the envelope and outflow. The size of the $^{34}$SO$_2$ emission is slightly smaller with a deconvolved size of $\sim0.4''\times0.2''$ ($\sim120\times60$~au), but this likely originates from the lower signal-to-noise of this transition. These emitting areas are consistent with that of HDO $3_{3,1}-4_{2,2}$ ($E_{\rm up} = 335$~K) and complex organics such as methanol (CH$_3$OH) and methyl formate (CH$_3$OCHO, see Fig.~\ref{fig:ALMA_maps_COMs}) and with the compact emission in the MIRI-MRS data (i.e., $0.5''\times0.3''$ in diameter, see Fig.~\ref{fig:so2_miri_map}).

The pure rotational lines of SO$_2$ and $^{34}$SO$_2$ detected by ALMA were fitted using the same LTE slab model fitting procedure as for the MIRI-MRS data (Sect.~\ref{subsec:slab_model_fitting}). The submillimeter line lists of both SO$_2$ and $^{34}$SO$_2$ were taken from the Cologne Database for Molecular Spectroscopy\footnote{\url{https://cdms.astro.uni-koeln.de/}} \citep[CDMS;][]{Muller2001,Muller2005_CDMS,Endres2016}, where the entry of SO$_2$ was mostly taken from \citet{Muller2005_SO2} and the entry of $^{34}$SO$_2$ is based on several spectroscopic works \citep[e.g.,][for transitions at ALMA Band~7 frequencies]{Lovas1985,Belov1998}. 
The ALMA data cover in total five pure rotational (i.e., $\nu=0$) transitions of SO$_2$ and also five transitions of the $^{34}$SO$_2$ isotopologue, see Table~\ref{tab:SO2_transitions}. However, several of these transitions are blended with strong emission of complex organics such as CH$_3$OCHO and CH$_3$CHO. Fortunately, two transitions of SO$_2$ and one transition of $^{34}$SO$_2$ are (relatively) unblended. This is important to determine the column density and excitation temperature. 
All lines of SO$_2$ and $^{34}$SO$_2$ are spectrally resolved with a FWHM of $3.5$~km~s$^{-1}$, which is similar to the FWHM of the lines of HDO $3_{3,1}-4_{2,2}$ and the complex organics.
No vibrational correction is necessary for the column densities derived from the pure rotational lines since the first excited vibrational states are included in the partition function.

Interestingly, the excitation temperature $T_{\rm ex}$ (i.e., rotational temperature $T_{\rm rot}$ since it is derived from pure rotational lines) derived for SO$_2$ from the ALMA data is very similar to that derived for MIRI: $104\pm5$~K, see Figs.~\ref{fig:ALMA_spectra} and \ref{fig:chi2_map_ALMA}. This suggests that ALMA is probing the same gas as that being probed with MIRI with the temperature reflecting the distribution of rotational levels within the $\nu=0$ state. The column density can be accurately constrained at $N_{\rm SO_2,ALMA} = 3.4\pm 1.1 \times 10^{16}$~cm$^{-2}$ for an emitting area with a radius of $R=85$~au. In contrast to MIRI, the column density and emitting area can be fairly accurately constrained since the lines are marginally optically thick ($\tau \sim 0.1$). Moreover, the derived column density does not suffer severely from optical depth effects given that the column density derived from the optically thin $^{34}$SO$_2$ isotopologue is $2.7\pm 0.5 \times 10^{15}$~cm$^{-2}$, giving a $^{32}$SO$_2$/$^{34}$SO$_2$ ratio of $13\pm5$ which is within a factor of 2 of the average $^{32}$S/$^{34}$S derived for the local ISM \citep[$^{32}$S/$^{34}$S = 22;][]{Wilson1999}. However, the total number of SO$_2$ molecules measured by ALMA is $\mathcal{N} = 1.6 \pm 0.2 \times 10^{47}$~molecules, which is $\sim4$ orders of magnitude lower than what is detected by MIRI. This discrepancy cannot be explained by a difference in dust opacity, given that such an effect would only worsen the discrepancy in the detected number of molecules between ALMA and MIRI. The most logical explanation is the importance of non-LTE effects for the ro-vibrational transitions detected by MIRI.

\section{Discussion}
\label{sec:discussion}
\subsection{Importance of non-LTE effects}
\label{subsec:non-LTE}
\subsubsection{Absence of the $\nu_1$ and $\nu_2$ bands}
\label{subsec:nu1_nu2}
If the emission originating from the $\nu_3$ asymmetric stretching mode were in LTE, both the $\nu_1$ symmetric stretching mode ($\lambda\sim8.5-9$~\mum) and $\nu_2$ bending mode ($\lambda\sim18-20$~\mum) should also have been detected. However, no SO$_2$ emission is detected in the $\nu_1$ and $\nu_2$ bands, see Fig.~\ref{fig:spec_and_cont_images}. The $\nu_2$ bending mode in particular has a significantly lower vibrational energy (518~cm$^{-1}$, 745~K) than the $\nu_3$ asymmetrical stretching mode \citep[1362~cm$^{-1}$, 1960~K;][]{Briggs1970,Person1982} and is therefore more easily collisionally excited.
The expected SO$_2$ flux in the $18-20$~\mum region by employing the best-fit LTE models of the $\nu_3$ band around 7.35~\mum (Sect.~\ref{subsec:res_MIRI}) and the vibrational ground state (i.e., $(\nu_1,\nu_2,\nu_3)=(0,0,0)$, denoted as $\nu=0$) in the ALMA data (Sect.~\ref{subsec:comp_to_ALMA}) are presented in Fig.~\ref{fig:so2_19mum}.
Similar to the analysis of the $\nu_3$ band, both models are corrected for an extinction of $A_{\rm V} = 55$~mag \citep[][]{Rochasubm} using a modified version of the \citet{McClure2009} extinction law (see Appendix~\ref{app:ext_corr}).
No obvious SO$_2$ emission features are present in the data whereas the best-fit $\nu_3$ LTE model clearly predicts that we should have seen the SO$_2$ at these wavelengths. On the other hand, the LTE model derived from the pure rotational lines in the ALMA data agrees with the non-detection of the $\nu_2$ band, predominantly due to the larger extinction at MIR wavelengths compared to mm-wavelengths \citep[e.g.,][]{McClure2009,Chapman2009}.


One solution for the lack of emission in the $\nu_2$ band could be that the line-to-continuum ratio is too low in this region, since the continuum flux level is on the order of $\sim1$~Jy at $18-20$~\mum compared to $\sim20$~mJy around 7.35~\mum. 
The predicted line emission peaks at $\sim1$~Jy for the strongest peaks of the $\nu_3$ model, corresponding to a line-to-continuum ratio of $1$. This suggests that this cannot explain the absence of the $\nu_2$ band since a line-to-continuum ratio down to 0.01 should still be detectable on a strong continuum. 
Moreover, this would also not explain the discrepancy between the number of molecules needed in the LTE slab models to explain the emission in $\nu=0$ state with ALMA and the $\nu_3$ band with MIRI. Scattered continuum radiation is present out to $\sim8$~\mum but not to longer wavelengths (see Fig.~\ref{fig:spec_and_cont_images}), suggesting that the $\nu_3$ band could be infrared pumped on $\sim100$~au scales whereas this does not occur for the $\nu_2$ band at 19~\mum. The infrared radiation around 19~\mum originates from thermal dust emission which is not extended (see Fig.~\ref{fig:spec_and_cont_images}).

Similarly to the $\nu_2$ bending mode, the $\nu_1$ symmetrical stretching mode between 8.5 and 9~\mum is overproduced by the best-fit $\nu_3$ model, see Fig.~\ref{fig:so2_9mum}. The absence of the emission from the $\nu_1$ band can be simply explained by the large extinction due to strong silicate absorption feature around these wavelengths 

\subsubsection{Infrared pumping}
The critical densities of ro-vibrational transitions are typically \mbox{$>10^{10}$}~cm$^{-3}$ \citep[e.g., for HCN and CO$_2$;][for SO$_2$ no collisional rates are available for the MIR transitions but they are expected to be similar]{Bruderer2015,Bosman2017}, suggesting that LTE conditions are only valid for ro-vibrational transitions in the inner $\lesssim1$~au. However, SO$_2$ is present in IRAS2A on $\sim100$~au scales (see Figs.~\ref{fig:so2_miri_map} and \ref{fig:ALMA_maps}) where the densities are typically $10^{6}-10^{8}$~cm$^{-3}$ \citep[e.g.,][]{Jorgensen2004,Jorgensen2022,Kristensen2012}, far below the critical densities of the MIR transitions, further suggesting that the vibrational level populations are not collisionally excited. The critical densities for the $\nu=0$ transitions probed by ALMA are much lower, on the order of $\sim10^{6}-10^{7}$~cm$^{-3}$, meaning that the LTE assumption is valid for these pure rotational lines. 

Despite the fact that the assumption of LTE is not valid for the MIR transitions, physical information can still be derived. The rotational temperature derived from the LTE models of the MIR $\nu_3$ transitions is well constrained given the accuracy of the fit to the $R$ and $Q$-branches. It is very similar to the rotational temperature derived for the $\nu=0$ state, for which the LTE assumption is valid.
Combined with the similar extent of the emission in the MIRI and ALMA images (Figs.~\ref{fig:so2_miri_map} and \ref{fig:ALMA_maps}, respectively), this strongly suggests that they are probing SO$_2$ gas on similar scales (i.e., $50-100$~au).
Because the rotational levels withing the $\nu=0$ state can be characterized by a single temperature, the column density of SO$_2$ derived from the ALMA data is also robust.
However, the total number of SO$_2$ molecules derived from the $\nu_3$ MIR transitions is found to be $>4$ orders of magnitude higher than what is measured for the $\nu=0$ state. This suggests that the $\nu_3$ vibrational state is more highly populated than just through collisional excitation by itself. 

A likely explanation for populating the $\nu_3$ vibrational level could be infrared pumping. Vibrational levels can become more highly populated due to the absorption of infrared photons in the presence of a strong infrared radiation field, boosting the line fluxes far above that expected from collisional excitation alone \citep[e.g.,][]{Bruderer2015,Bosman2017}. During the infrared radiative pumping process, the distribution of rotational levels within the vibrational states is largely maintained since only $\Delta J =0,\pm1$ transitions are allowed. Infrared pumping thus leads to a more highly populated vibrational level without significantly changing the rotational distribution (i.e., rotational temperature), supporting the similarity in rotational temperatures measured in the $\nu_3=1$ and $\nu=0$ states.

The importance of infrared pumping can be further quantified by comparing the results of the LTE models to the $\nu_3$ band and the $\nu=0$ state. Assuming the $\nu_3=1$ vibrational level is radiatively pumped from the $\nu=0$ state, the vibrational level population is set following the Boltzmann distribution by a vibrational temperature $T_{\rm vib}$ that is different from the rotational temperature, $T_{\rm vib} \neq T_{\rm rot}$. The difference between the total number of molecules predicted in LTE models of the MIRI and ALMA data can then be approximated as,
\begin{align}
    \frac{\mathcal{N}_{\rm SO_2,MIRI}}{\mathcal{N}_{\rm SO_2,ALMA}} \propto \frac{e^{-h \nu/(k_{\rm B} T_{\rm vib})}}{e^{-h \nu/(k_{\rm B} T_{\rm rot})}},
    \label{eq:T_vib_Trot}
\end{align}
where $\mathcal{N}_{\rm ALMA}$ and $\mathcal{N}_{\rm MIRI}$ are the total number of molecules needed in the LTE models of the ALMA and MIRI data, respectively, $T_{\rm rot}$ the derived rotational temperature in the $\nu=0$ state with ALMA (i.e., $104\pm5$~K; Sect.~\ref{subsec:comp_to_ALMA}), $T_{\rm vib}$ the vibrational temperature, $\nu$ the frequency of the MIR transitions, and $h$ and $k_{\rm B}$ Planck's and Boltzmann's constants, respectively. In Eq.~\eqref{eq:T_vib_Trot}, $g$ factors have been neglected for simplicity. In the case where both $\nu_3=1$ and $\nu=0$ are collisionally excited, $T_{\rm vib}=T_{\rm rot}$ and the number of SO$_2$ molecules predicted by the LTE models of the MIRI and ALMA data should have been equal, $\mathcal{N}_{\rm ALMA} = \mathcal{N}_{\rm MIRI}$. However, for the derived $\frac{\mathcal{N}_{\rm SO_2,MIRI}}{\mathcal{N}_{\rm SO_2,ALMA}} \sim 2\times10^4$ and $T_{\rm rot} = 104$~K, Eq.~\eqref{eq:T_vib_Trot} results in $T_{\rm vib} \sim 200$~K at a wavelength of $7.35$~\mum (i.e., $\nu = 40.79$~THz), which is $\sim80$~K higher than the derived $T_{\rm rot}$ of the vibrational ground state. We estimate an uncertainty on $T_{\rm vib}$ of 50~K based on the assumption of similar rotational temperatures in the $\nu_3=1$ and $\nu=0$ states and ignoring $g$ factors in Eq.~\eqref{eq:T_vib_Trot}.

If infrared pumping is indeed the cause for the elevated vibrational temperature, the brightness temperature of the infrared radiation around 7.35~\mum, $T_{\rm IR}$, has to be similar to $T_{\rm vib}$,
\begin{align}
    T_{\rm IR} = \frac{h \nu}{k_{\rm B}} \ln^{-1}\left(1 + \frac{2 h \nu^3}{I_\nu c^2}\right),
\end{align}
where $I_\nu$ is the observed intensity (in Jy sr$^{-1}$) of the continuum at a frequency $\nu$ and $c$ is the speed of light. Given that $I_\nu \sim 6\times10^4$~MJy~sr$^{-1}$ (i.e., 1.4~Jy~arcsec$^{-2}$) around 7.35~\mum (i.e., $40.79$~THz) and assuming an extinction of $A_{\rm V} = 55$~mag \citep[][]{Rochasubm}, this results in $T_{\rm IR} \sim 180$~K. 
This is in very good agreement with $T_{\rm vib} \sim 200$~K and therefore strongly suggests that the vibrational temperature of the $\nu_3$ band of SO$_2$ is indeed set by the infrared radiation field of the protostar rather than by collisions. 

In turn, this means that both MIRI and ALMA are in fact tracing the same molecular gas with the same rotational temperature, but that the transitions in the $\nu_3$ band are just pumped by the strong infrared radiation field of the central star and originate from a region much smaller than 5000~au. ALMA thus measures the true number of SO$_2$ molecules ($1.8\pm0.2\times10^{47}$), and in order to directly derive the physical properties (i.e., excitation temperature, column density, emitting area) from the ro-vibrational transition of SO$_2$ at MIR wavelengths it is important to take into account non-LTE processes such as infrared radiative pumping.

\subsection{Physical origin of the SO$_2$}
\label{subsec:origin_SO2}
The SO$_2$ emission detected with both ALMA and MIRI is clearly tracing warm inner regions of the protostellar system on disk scales of $\lesssim100$~au in radius. This implies that the emission either originates from the central hot core where many complex organics are detected \citep[e.g.,][]{Jorgensen2005,Bottinelli2007,Maury2014,Taquet2015}, or that the emission originates from an accretion shock at the disk-envelope interface \citep[e.g.,][]{ArturdelaVillarmois2019,ArturdelaVillarmois2022,vanGelder2021}. Alternatively, the SO$_2$ could be located in a jet or outflow close to the source \citep[e.g.,][]{Codella2014,Taquet2020,Tychoniec2021}, or even in a disk wind \citep[e.g.,][]{Tabone2017}. However, since the emission in the MIRI data is clearly centrally peaked with only little extended emission ($\lesssim100$~au in radius; see Fig.~\ref{fig:so2_miri_map}) and the line widths of the pure rotational lines detected by ALMA are $\sim3.5$~km~s$^{-1}$ \citep[compared to $>10$~km~s$^{-1}$ typically observed in outflows; e.g.,][]{Taquet2020,Tychoniec2019,Tychoniec2021}, an extended outflow origin is less likely. On the contrary, CO$_2$ is clearly detected in the outflow and not on the continuum peak (see Fig.~\ref{fig:co2_miri_map}), which will be further discussed in a separate paper. 

One way to investigate further the possible origin of the SO$_2$ emission is to estimate its abundance with respect to H$_2$. The most direct way to determine the number of H$_2$ molecules is using the detected H$_2$ MIR lines, see Appendix~\ref{app:H2_analysis}. 
However, H$_2$ does not only trace the warm inner regions but is also present in outflows and disk winds (Tychoniec et al. in prep.). Indeed, in IRAS2A H$_2$ is mostly located in the outflow toward the southwest (see Fig.~\ref{fig:co2_miri_map}).
Moreover, the rotational temperature of the warm component is \mbox{$>300$}~K (see Fig.~\ref{fig:H2_rot_diag}), which further suggests that H$_2$ may not be tracing the same gas as SO$_2$ ($T_{\rm rot} = 104\pm5$~K). 

Another method for determining the amount of $\sim100$~K H$_2$ gas is to use the ALMA Band~7 continuum. The total gas mass, $M_{\rm gas}$, can be determined using the equation from \citet{Hildebrand1983},
\begin{align}
    M_{\rm gas} = 100 \frac{F_{\rm \nu} d^2}{\kappa_{\rm \nu} B_{\rm \nu}(T_{\rm dust})},
\end{align}
where $F_{\rm \nu}$ is the continuum flux density at a frequency $\nu$, $d$ the distance \citep[293~pc;][]{Ortiz-Leon2018}, $\kappa_{\rm \nu}$ the dust opacity,  $B_{\rm \nu}(T_{\rm dust})$ the Planck function evaluated at a dust temperature $T_{\rm dust}$, and the factor 100 the assumed gas-to-dust mass ratio. Here, $T_{\rm dust}$ is assumed to be 30~K, which is a typical dust temperature for protostellar envelopes at $\sim100$~au scales \citep{Whitney2003}. For a measured $F_{\rm \nu}=0.396\pm0.079$~Jy in a $1.4''$ diameter aperture (i.e., the same as was used in the spectral extraction) and a dust opacity of $\kappa_{\rm \nu} = 1.84$~cm$^{-2}$~g$^{-1}$ at a frequency of 340~GHz \citep{Ossenkopf1994}, a gas mass of $M_{\rm gas} = 0.11\pm0.02$~M$_\odot$ is derived. The derived gas mass is consistent within a factor of two with other recent measurements \citep[e.g.,][]{Tobin2018,Tychoniec2020}. Assuming that this gas mass is predominantly in H$_2$ \citep[i.e. taking a mean molecular weight $\mu_{\rm H_2} = 2.8$ per hydrogren atom for gas composef of 71\% hydrogen;][]{Kauffmann2008}, this results in $\mathcal{N_{\rm H_2}} = 4.7 \pm 0.9 \times 10^{55}$~molecules. This is in good agreement with $\mathcal{N_{\rm H_2}} = 6.6 \pm 0.9 \times 10^{55}$~molecules derived for IRAS2A from C$^{18}$O lines up to $J=9-8$ with {\it Herschel}-HIFI \citep[assuming a CO/H$_2$ ratio of 10$^{-4}$ and rotational temperature of $\sim40$~K;][]{Yildiz2013}. However, the derived radius of the SO$_2$ emitting area ($\sim100$~au) is smaller than the physical radius of the aperture (205~au for $1.4''$ diameter aperture) used for computing $\mathcal{N_{\rm H_2}}$. The derived $\mathcal{N_{\rm H_2}}$ can be scaled to the derived emitting area of SO$_2$ assuming that the density scales as $n_{\rm H} \propto R^{-p}$,
\begin{align}
    \mathcal{N_{\rm H_2,T>100K}} = \mathcal{N_{\rm H_2}} \left(\frac{R_{\rm 100K}}{R_{\rm ap}}\right)^{3-p},
\end{align}
where $R_{\rm 100K}$ is the derived size emitting area of SO$_2$ from the ALMA and MIRI-MRS maps ($\sim100$~au, see Sect.~\ref{sec:results}) and $R_{\rm ap}$ is physical radius of the aperture (205~au for $1.4''$ diameter aperture). Taking a density powerlaw index of 1.7 for IRAS2A1 \citep{Kristensen2012}, this results in $\mathcal{N_{\rm H_2,T>100K}} = 1.8\pm0.4 \times 10^{55}$~molecules. 
This is almost equal to what is derived directly from the H$_2$ lines ($1.7\pm0.9\times10^{55}$ molecules; Appendix~\ref{app:H2_analysis}) which suggests that the H$_2$ 0-0 lines themselves are likely sensitive to most of the $T>100$~K gas despite the high rotational temperature ($T_{\rm rot} = 356\pm41$~K).

The total number of SO$_2$ molecules derived from the emission of the vibrational ground state in the ALMA data can be directly compared to H$_2$. This results in an SO$_2$ abundance of the order of $1.0\pm0.3\times10^{-8}$ with respect to H$_2$, which is in agreement with other recent SO$_2$ abundance measurements in low-mass Class~0 systems \citep[$>6.6\times10^{-10}$;][]{ArturdelaVillarmois2023}. The abundance of $\sim10^{-8}$ implies that the SO$_2$ gas is not the dominant sulfur carrier in IRAS2A since the cosmic [S/H] abundance is about $\sim10^{-5}$ \citep{Savage1996,Goicoechea2006}. It is also means that SO$_2$ does not contain a significant amount of the total volatile sulfur budget in dense clouds \citep[volatile {[S/H]}$\sim 10^{-7}$, i.e., the amount of sulfur that is not locked up in refractory formats;][]{Woods2015,Kama2019}. Furthermore, the derived SO$_2$ abundance is on the lower side compared with estimates of the SO$_2$ abundances in interstellar ices \citep[$10^{-8}-10^{-7}$;][]{Boogert1997,Boogert2015,Zasowski2009} and cometary ices \citep[$\sim10^{-7}$;][]{Altwegg2019,Rubin2019}. In particular, SO$_2$ ice was also recently detected in absorption toward IRAS2A itself with a very similar abundance with respect to H$_2$ as other low-mass protostars \citep[$\sim10^{-7}$;][]{Rochasubm}. 

Although the gaseous SO$_2$ abundance is on the lower side compared to the ices, it suggests that the observed gaseous SO$_2$ could be sublimated from the ices in the central hot core since the sublimation temperature of SO$_2$ \citep[$\sim60$~K, $E_{\rm bin} = 3010$~K;][]{Penteado2017} is similar to H$_2$O and many complex organics. It also naturally explains the compactness of the SO$_2$ emission in both the MIRI data and the ALMA data (see Figs.~\ref{fig:so2_miri_map} and \ref{fig:ALMA_maps}) with an extent very similar to that of H$_2$O, HDO, and complex organics (see Figs.~\ref{fig:co2_miri_map} and \ref{fig:ALMA_maps_COMs}). In fact, the emitting areas derived from the MIRI-MRS and ALMA integrated intensity maps ($\sim100$~au in radius) agree well with that of a hot core based on the luminosity \citep[i.e., $R_{\rm 100 K} = 147$~au, where $R_{\rm 100 K} \approx 15.4\sqrt{L_{\rm bol}/L_\odot}$;][]{Bisschop2007,van'tHoff2022}.
The hot core origin is further supported by the line width in the ALMA data ($\sim3.5$~km~s$^{-1}$) which is similar to that of lines from HDO and complex organics. 

Another possibility for the SO$_2$ emission could be weak shocks on $\lesssim100$~au scales such as accretion shocks at the disk-envelope boundary \citep[e.g.,][]{Sakai2014,Oya2019,ArturdelaVillarmois2019,ArturdelaVillarmois2022}. Recent shock models have suggested that SO$_2$ could be a good tracer of such accretion shocks \citep{Miura2017,vanGelder2021}. Abundances up to $10^{-8}-10^{-7}$ with respect to H$_2$ are easily reached in low-velocity accretion shocks, as long as a significant ultraviolet (UV) radiation field is present \citep[i.e., stronger than the interstellar radiation field;][]{vanGelder2021}. Given the luminosity of IRAS2A \citep[$\sim60-90$~L$_\odot$;][]{Murillo2016,Karska2018}, having a considerable UV field in the inner envelope out to $100$~au scales is likely \citep[see e.g.,][]{vanKempen2009,Yildiz2012,Yildiz2015}. 
The emission morphology seen in the ALMA data (Fig.~\ref{fig:ALMA_maps}) shows that the emission is extended more in the direction of the outflow than along the disk. Furthermore, the line width of $\sim3.5$~km~s$^{-1}$ is consistent with those of the complex organics whereas accretion shocks are expected to show broader emission lines \citep[$\sim10$~km~$^{-1}$][]{Oya2019,ArturdelaVillarmois2019,ArturdelaVillarmois2022}. A hot core origin is therefore a more likely explanation for the gaseous SO$_2$.

The SO$_2$ emission in IRAS2A could be tracing similar components as the MIR SO$_2$ absorption that was detected toward multiple high-mass protostellar sources \citep[e.g,][]{Keane2001,Dungee2018,Nickerson2023}. Typical temperatures of this SO$_2$ absorption are $\sim100-300$~K \citep[e.g.,][]{Dungee2018,Nickerson2023}, which is very similar to the rotational temperature derived for IRAS2A ($\sim100$~K), though temperatures up to $\sim700$~K have also been reported \citep{Keane2001}. The most common origin of the SO$_2$ absorption toward these high-mass sources is also suggested to be a hot core rather than a shock, similar to what SO$_2$ is tracing in IRAS2A, with the main difference that it is in absorption against the bright infrared continuum of the central high-mass protostar. The typical SO$_2$ abundances derived for these high-mass protostars are $\gtrsim10^{-7}$ with respect to H$_2$ \citep[][]{Keane2001,Dungee2018,Nickerson2023}, which is an order of magnitude higher than what is derived for IRAS2A. 

It remains unknown whether the presence of ro-vibrational lines of SO$_2$ in IRAS2A is unique or whether it is more common among low-mass protostars.
One other low-mass protostar in the JOYS+ sample, NGC~1333~IRAS1A, shows emission of the SO$_2$ $\nu_3$ $Q$-branch ($R$- and $P$-branches are not detected) and similarly does not show any emission from the $\nu_1$ and $\nu_2$ bands, suggesting that infrared radiative pumping may also be responsible for the emission in the $\nu_3$ band of this source.
IRAS2A has a high luminosity \citep[$\sim60-90$~L$_\odot$;][]{Murillo2016,Karska2018} and is suggested to currently be in a burst phase \citep[e.g.,][]{Hsieh2019,van'tHoff2022}, whereas IRAS1A has a lower luminosity \citep[$\sim10$~L$_\odot$][]{Tobin2016}. On the other hand, the low-mass Class~0 protostar IRAS~15398-3359 does not show gaseous SO$_2$ and has a lower luminosity of $\sim1.5$~L$_\odot$ \citep{Yang2018,Yang2022}. 
Given the importance of infrared pumping in detecting the SO$_2$ lines, a high luminosity could be important. However, a larger sample of JWST/MIRI-MRS spectra of low-mass and high-mass embedded protostellar systems is needed to further investigate the importance of luminosity (and other source properties) on the presence of MIR SO$_2$ emission. Future MIRI-MRS observations from the JOYS+ program will cover a few other luminous low-mass protostellar systems \citep[e.g., Serpens~SMM1, $L_{\rm bol} \sim 100$~L$_\odot$][]{Karska2018}, as well as several high-mass protostellar systems \citep[e.g., IRAS 18089-1732, $L_{\rm bol} \sim 10^4$~L$_\odot$;][]{Urquhart2018}.


\section{Conclusions}
\label{sec:conclusions}
This paper presents one of the first medium-resolution mid-infrared spectra and images taken with JWST/MIRI-MRS of a Class~0 protostellar source, NGC~1333~IRAS2A, and shows the first detection of gas-phase SO$_2$ emission at MIR wavelengths.
The spectral lines of the $\nu_3$ asymmetric stretching mode of SO$_2$ are analyzed using LTE slab models, giving a rotational temperature of $92\pm8$~K. This is very similar to the rotational temperature of $104\pm5$~K derived from the pure rotational lines in the high-resolution ALMA data. Since the SO$_2$ emission in the MIRI-MRS data is optically thin, the column density could not be constrained accurately due to the degeneracy with the emitting area. However, the total number of molecules can be constrained and is predicted by LTE models of the MIRI-MRS data to be a factor of $\sim2\times10^4$ higher than derived from the ALMA data. Based on these results, our main conclusions are as follows:

\begin{itemize}
    \item The $\nu_3$ asymmetric stretching mode of SO$_2$ detected around 7.35~\mum with MIRI is not in LTE but rather radiatively pumped by a strong infrared radiation field scattered out to $\sim100$~au distances. 
    The discrepancy between the best-fit LTE models of the MIRI-MRS and ALMA data can be explained by a higher vibrational temperature ($\sim200$~K) than the rotational temperature derived within the vibrational ground state ($104\pm5$~K). The vibrational temperature is consistent with the brightness temperature of the continuum around 7.35~\mum ($\sim180$~K). The similarity in rotational temperatures suggests that MIRI-MRS and ALMA are in fact still tracing the same molecular gas.
    \item Assuming that ALMA is probing the total amount of SO$_2$, the abundance of gaseous SO$_2$ is estimated to be $1.0\pm0.3\times10^{-8}$ with respect to H$_2$, which is consistent with both a hot core and accretion shock origin. Based on the size of the emitting area ($\sim100$~au in radius) and the small line width of the SO$_2$ lines ($\sim3.5$~km~s$^{-1}$) in the ALMA data, a hot core origin is suggested to be the most likely origin.
\end{itemize}

\noindent Our results show the importance of non-LTE effects in analyzing ro-vibrational lines at MIR wavelengths. 
The synergy between JWST probing the hot spots and the spatial and spectral resolution of ALMA proves to be crucial in determining the physical origin of molecules in embedded protostellar systems.
Future astrochemical modeling with non-LTE effects included will be valuable for inferring the physics and the chemistry of SO$_2$ in the earliest phases of star formation and their connection to the sulfur depletion problem.


\begin{acknowledgements}
We would like to thank the anonymous referee for their constructive comments on the manuscript, and Valentin Christiaens, Matthias Samland, and Danny Gasman for valuable support with the MIRI-MRS data reduction.
This work is based on observations made with the NASA/ESA/CSA James Webb Space Telescope. The data were obtained from the Mikulski Archive for Space Telescopes at the Space Telescope Science Institute, which is operated by the Association of Universities for Research in Astronomy, Inc., under NASA contract NAS 5-03127 for JWST. These observations are associated with program \#1236.
The following National and International Funding Agencies funded and supported the MIRI development: NASA; ESA; Belgian Science Policy Office (BELSPO); Centre Nationale d’Études Spatiales (CNES); Danish National Space Centre; Deutsches Zentrum fur Luftund Raumfahrt (DLR); Enterprise Ireland; Ministerio De Economiá y Competividad; The Netherlands Research School for Astronomy (NOVA); The Netherlands Organisation for Scientific Research (NWO); Science and Technology Facilities Council; Swiss Space Office; Swedish National Space Agency; and UK Space Agency. 

This paper makes use of the following ALMA data: ADS/JAO.ALMA\#2021.1.01578.S. ALMA is a partnership of ESO (representing its member states), NSF (USA) and NINS (Japan), together with NRC (Canada), MOST and ASIAA (Taiwan), and KASI (Republic of Korea), in cooperation with the Republic of Chile. The Joint ALMA Observatory is operated by ESO, AUI/NRAO and NAOJ.
The PI acknowledges assistance from Aida Ahmadi from Allegro, the European ALMA Regional Center node in the Netherlands.

MvG, EvD, LF, HL, KS, and WR acknowledge support from ERC Advanced grant 101019751 MOLDISK, TOP-1 grant 614.001.751 from the Dutch Research Council (NWO), The Netherlands Research School for Astronomy (NOVA), the Danish National Research Foundation through the Center of Excellence “InterCat” (DNRF150), and DFG-grant 325594231, FOR 2634/2.

The work of MER was carried out at the Jet Propulsion Laboratory, California Institute of Technology, under a contract with the National Aeronautics and Space Administration.

P.J.K. acknowledges support from the Science Foundation Ireland/Irish Research Council Pathway programme under Grant Number 21/PATH-S/9360.

LM acknowledges the financial support of DAE and DST-SERB research grants (SRG/2021/002116 and MTR/2021/000864) from the Government of India.

\end{acknowledgements}

\bibliographystyle{aa} 
\bibliography{refs} 

\onecolumn
\appendix
\section{Additional ALMA figures and table}
\label{app:ALMA_data}

\renewcommand{\arraystretch}{1.7}
\begin{table*}[h]
    \centering
    \caption{Transitions of SO$_2$ and $^{34}$SO$_2$ covered in the 2021.1.01578.S ALMA program. }\label{tab:SO2_transitions}
    \begin{tabular}{lrclcrcrc}
    \hline\hline
    Isotopologue & \multicolumn{3}{c}{Transition} & Frequency & $E_{\rm up}$ & $A_{\rm ij}$ &  $g_{\rm up}$ & Detection \\ \cline{2-4}
                 & $J_{K_{\rm a},K_{\rm c}}$ & - & $J'_{K_{\rm a}',K_{\rm c}'}$ & GHz & K & s$^{-1}$ & & \\
    \hline
    SO$_2$       & 8$_{2,6}$ & - & 7$_{1,7}$ & 334.673353 & 43.1 & $1.29\times10^{-4}$ & 17 & Y \\
                 & 23$_{3,21}$ & - & 23$_{2,22}$ & 336.089228 & 276.0 & $2.67\times10^{-4}$ & 47 & B  \\
                 & 16$_{7,9}$ & - & 17$_{6,12}$ & 336.669581 & 245.1 & $5.84\times10^{-5}$ & 33 & B  \\
                 & 13$_{2,12}$ & - & 12$_{1,11}$ & 345.338538 & 93.0 & $2.38\times10^{-4}$ & 27 & B  \\
                 & 16$_{4,12}$ & - & 16$_{3,13}$ & 346.523878 & 164.5 & $3.43\times10^{-4}$ & 33 & Y  \\
    \hline
    $^{34}$SO$_2$& 20$_{8,12}$ & - & 21$_{7,15}$ & 334.996573 & 334.0 &                     $6.19\times10^{-5}$ & 41 & N  \\
                 & 9$_{4,6}$ & - & 9$_{3,7}$ & 345.285620 & 79.2 & $2.88\times10^{-4}$ & 19 & B  \\
                 & 4$_{4,0}$ & - & 4$_{3,1}$ & 345.678787 & 47.1 & $1.31\times10^{-4}$ & 9 & B \\
                 & 17$_{4,14}$ & - & 17$_{3,15}$ & 345.929349 & 178.5 & $3.64\times10^{-4}$ & 35 & Y \\
                 & 28$_{2,2}$6 & - & 28$_{1,27}$ & 347.483124 & 390.7 & $2.65\times10^{-4}$ & 57 & N \\
    \hline
    \end{tabular}
    \tablefoot{The spectroscopic information of both SO$_2$ and $^{34}$SO$_2$ are taken from CDMS \citep[][]{Muller2001,Muller2005_CDMS,Endres2016}. The entry of SO$_2$ was mostly based on the work of \citet{Muller2005_CDMS} and the entry of $^{34}$SO$_2$ is based of several spectroscopic works \citep[e.g.,][for transitions at ALMA Band~7 frequencies]{Lovas1985,Belov1998}. The last column indicates whether the specific transition was detected (Y), blended (B), or not detected (N).}
\end{table*}
\renewcommand{\arraystretch}{1.0}


\begin{figure*}[h]
    \centering
    \includegraphics[width=\linewidth]{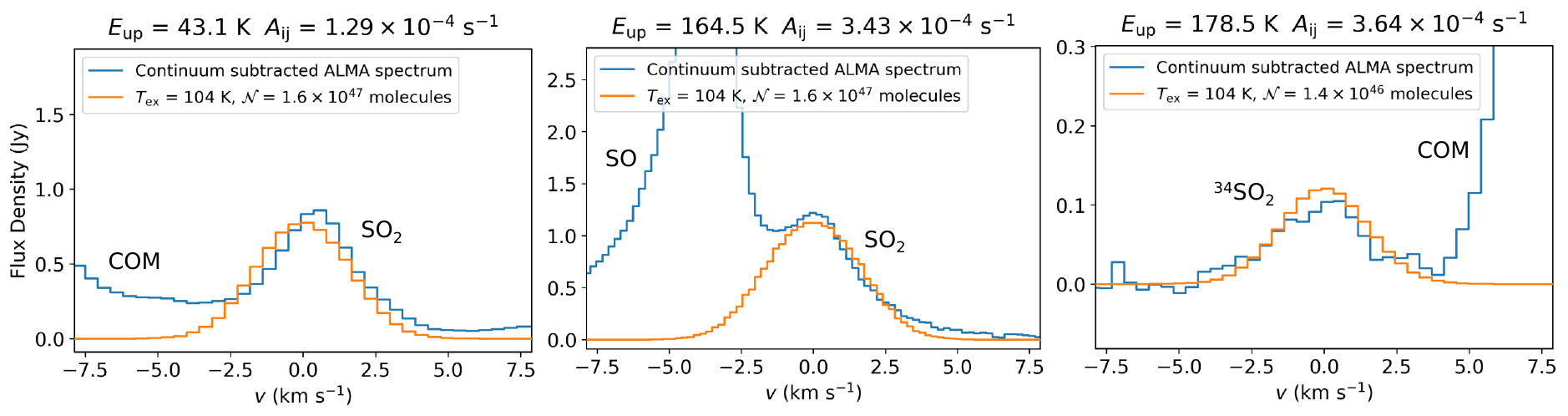}
    \caption{Continuum subtracted spectra (blue) and best-fit LTE slab models (orange) to the ALMA data for the SO$_2$ $8_{2,6}-7_{1,7}$ (left, $E_{\rm up} = 43$~K), SO$_2$ $16_{4,12}-16_{3,13}$ (middle, $E_{\rm up} = 164$~K), and $^{34}$SO$_2$ $17_{4,14}-17_{3,15}$ (right, $E_{\rm up} = 178$~K) transitions. All other transitions (see Table~\ref{tab:SO2_transitions}) of SO$_2$ and $^{34}$SO$_2$ are either blended with other molecular species or not detected. 
    } 
    \label{fig:ALMA_spectra}
\end{figure*}

\begin{figure}
    \centering
    \includegraphics[width=0.5\linewidth]{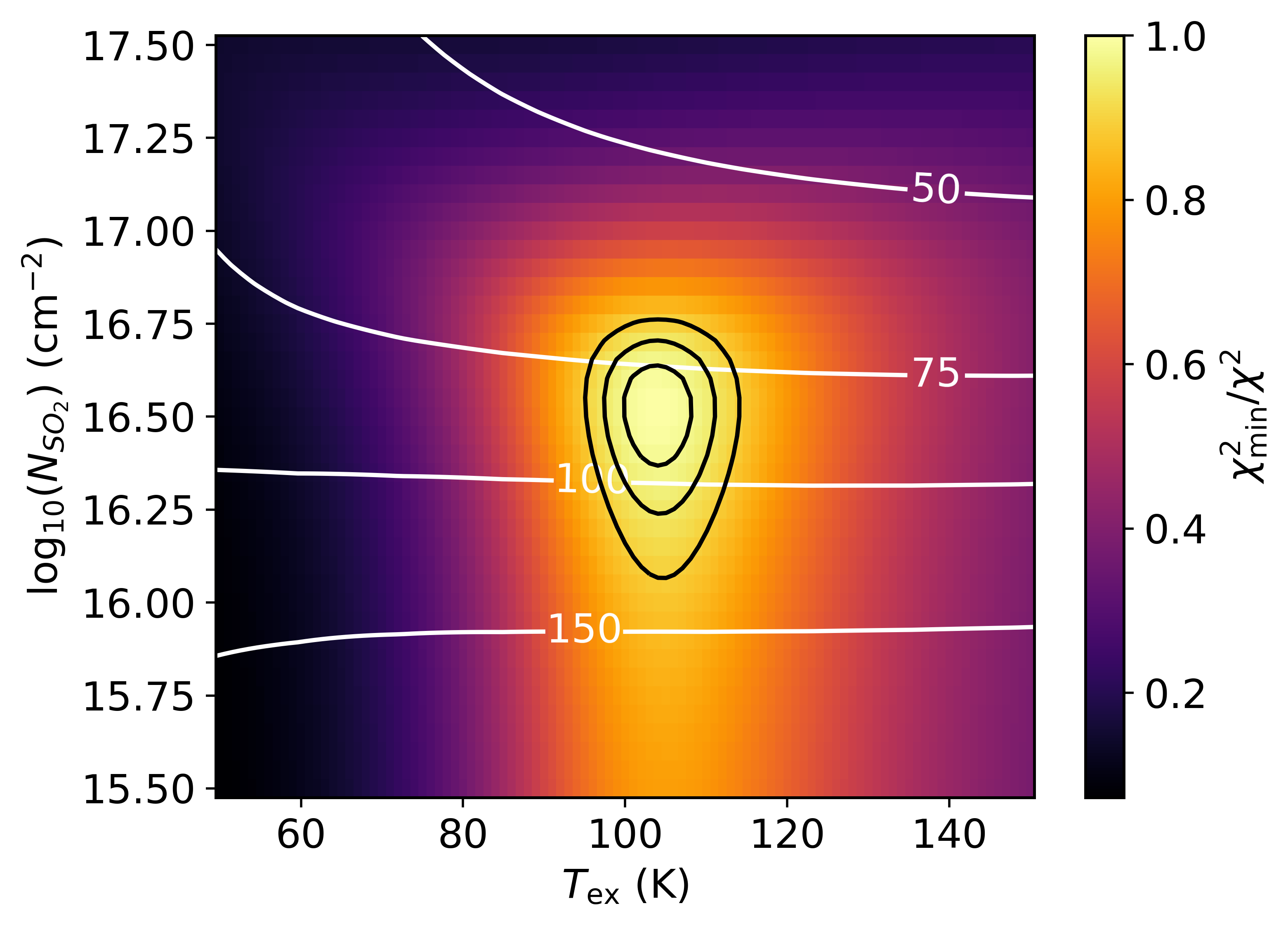}
    \caption{$\chi^2$ map for SO$_2$ derived from pure rotational lines in the ALMA data. The $\chi^2$ values are inversely normalized by the minimum $\chi^2$ and shown in color. The 1, 2, and 3$\sigma$ confidence intervals are presented as the black contours. The best-fitting emitting radius $R$ (in au) is represented by the white contours and is calculated for each grid point by minimizing the $\chi^2$ assuming a circular emitting area of $\pi R^2$.
    The $\chi^2$ map indicates that the SO$_2$ emission has an excitation temperature of $104\pm5$~K and is marginally optically thick ($\tau\sim0.1$) with an emitting area of $\sim85$~au in radius.} 
    \label{fig:chi2_map_ALMA}
\end{figure}

\begin{figure}
    \centering
    \includegraphics[width=0.33\linewidth]{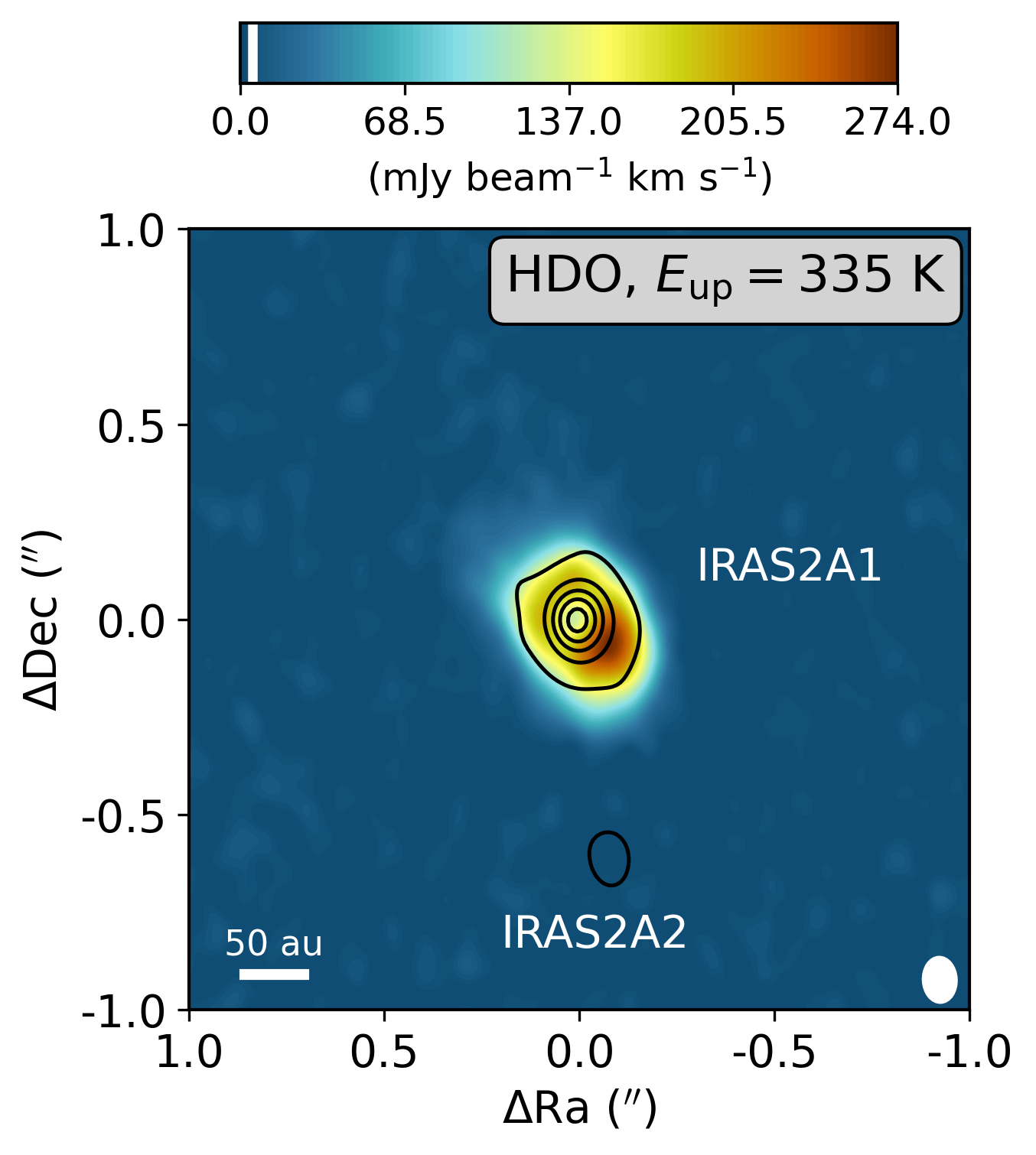}
    \includegraphics[width=0.33\linewidth]{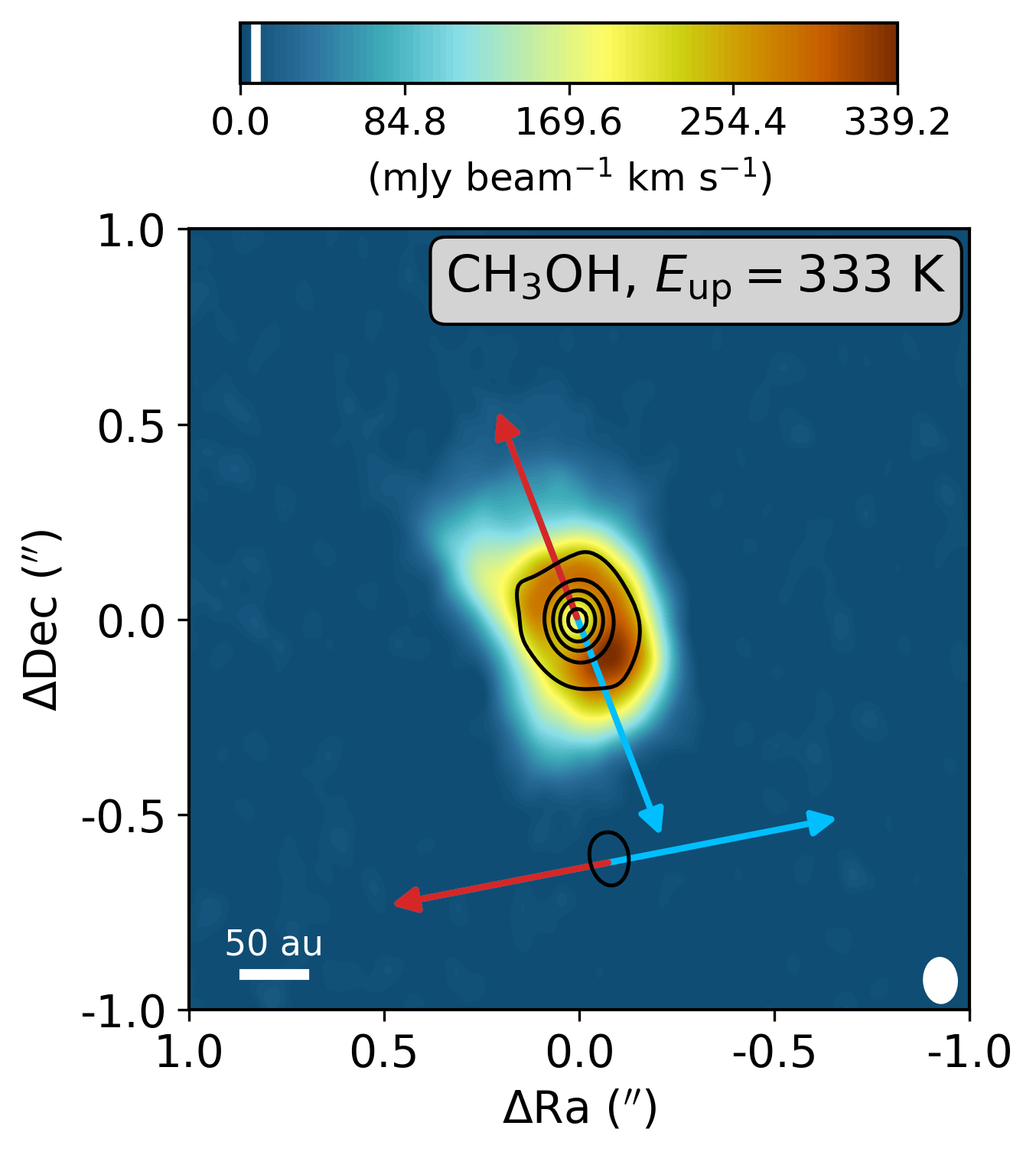}
    \includegraphics[width=0.33\linewidth]{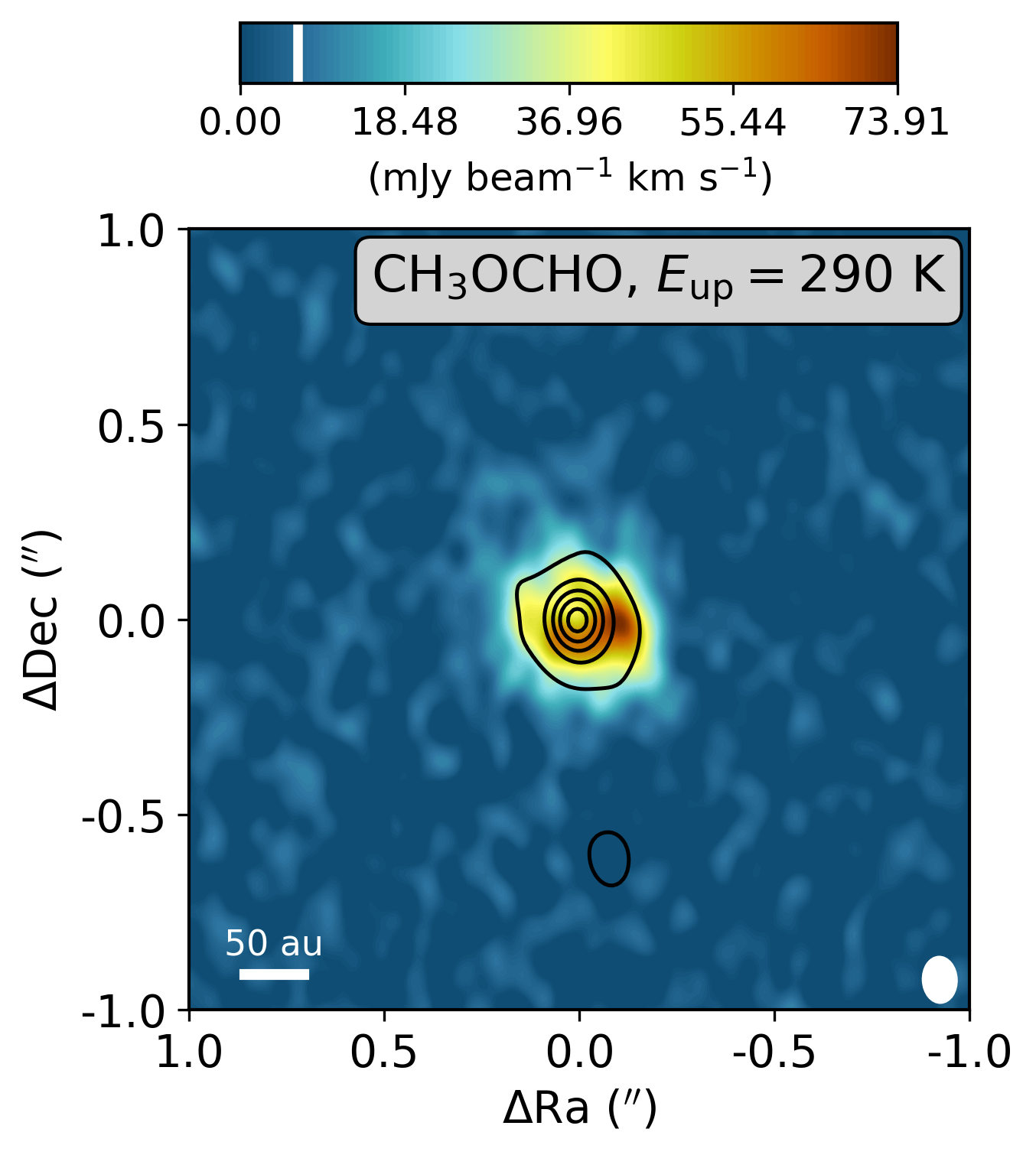}
    \caption{Integrated intensity maps of the HDO $3_{3,1}-4_{2,2}$ ($E_{\rm up} = 335$~K), CH$_3$OH $16_{1,15}-15_{2,14}$ (middle, $E_{\rm up} = 332$~K), and CH$_3$OCHO $27_{10,17}-26_{10,16}$ (right, $E_{\rm up} = 290$~K) transitions in color. The images are integrated over [-2,2]~km~s$^{-1}$ with respect to the $V_{\rm lsr}$ of 6.7~km~s$^{-1}$. The white vertical bar in the colorbar on top of each image indicates the 3$\sigma$ threshold. The 0.875~mm continuum is overlaid in the black contours. The main continuum peak is associated to IRAS2A1 and the secondary peak toward the south with IRAS2A2. The direction of the two outflows originating from IRAS2A1 and IRAS2A2 are indicated with the colored arrows in the middle panel \citep{Tobin2015_IRAS2A}. The size of the beam is shown in the bottom right and in the bottom left a scale bar is displayed.} 
    \label{fig:ALMA_maps_COMs}
\end{figure}

\clearpage
\section{Additional MIRI-MRS figures}

\begin{figure}[h]
    \centering
    \includegraphics[width=0.45\linewidth]{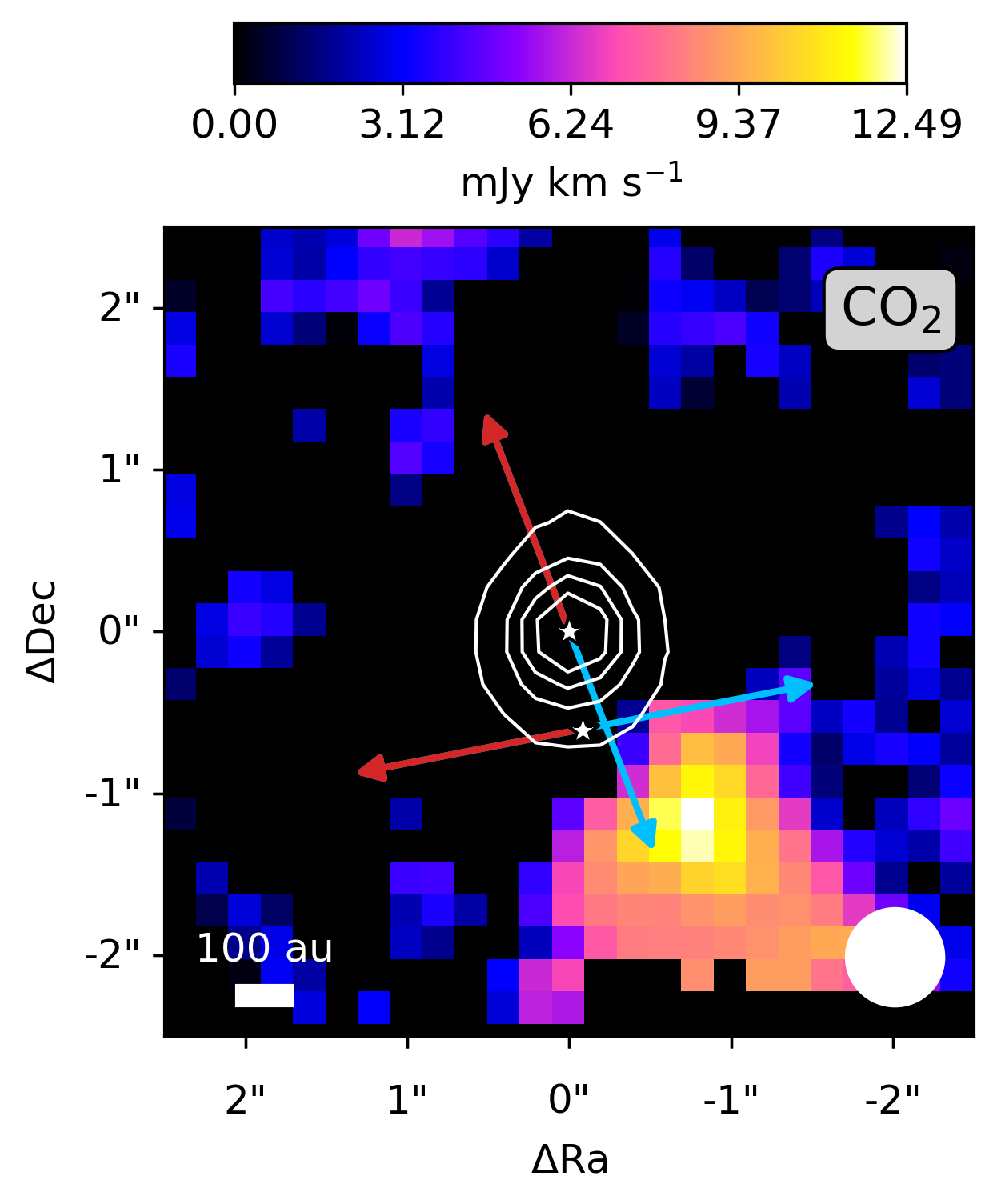}
    \includegraphics[width=0.45\linewidth]{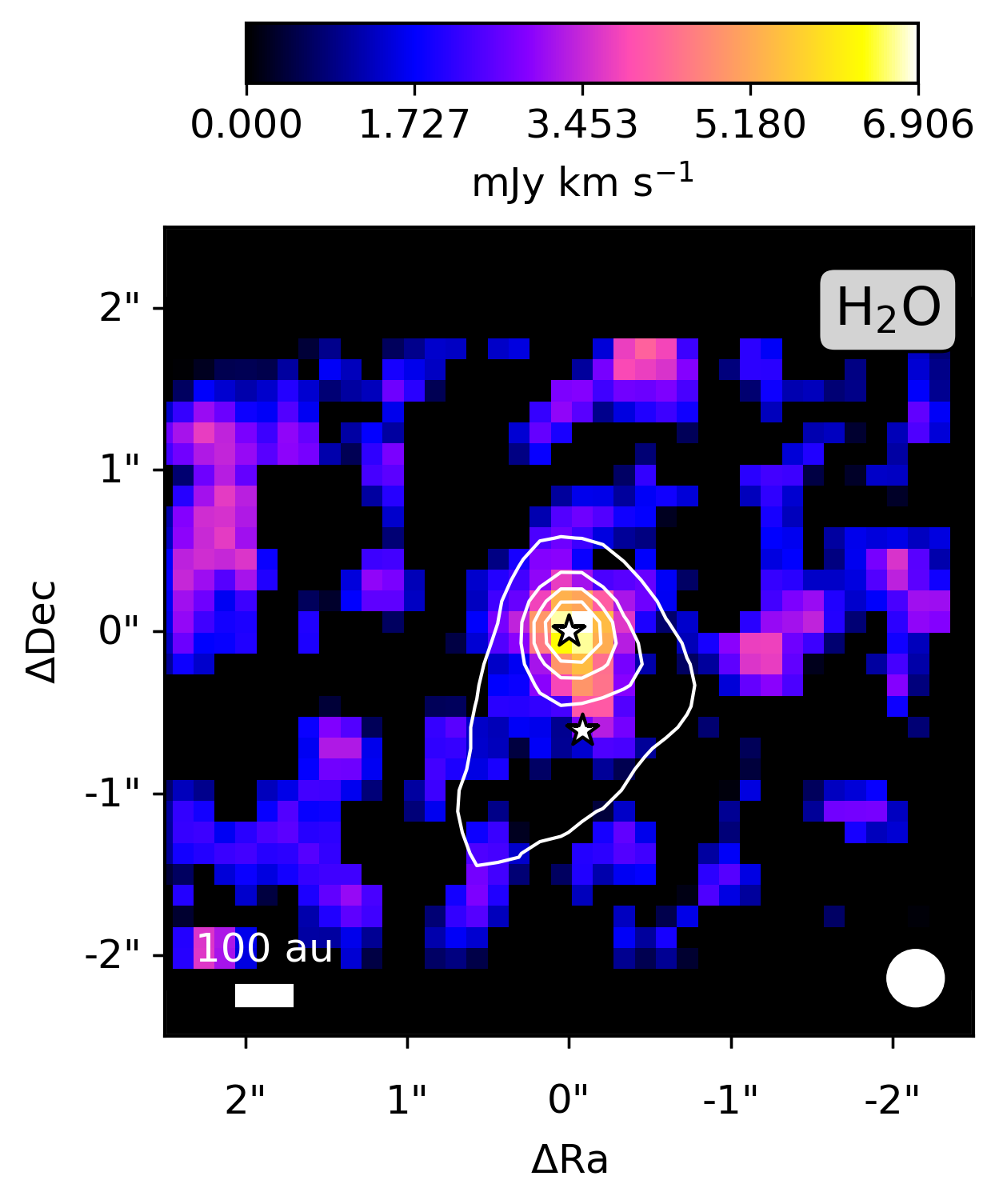}
    \includegraphics[width=0.45\linewidth]{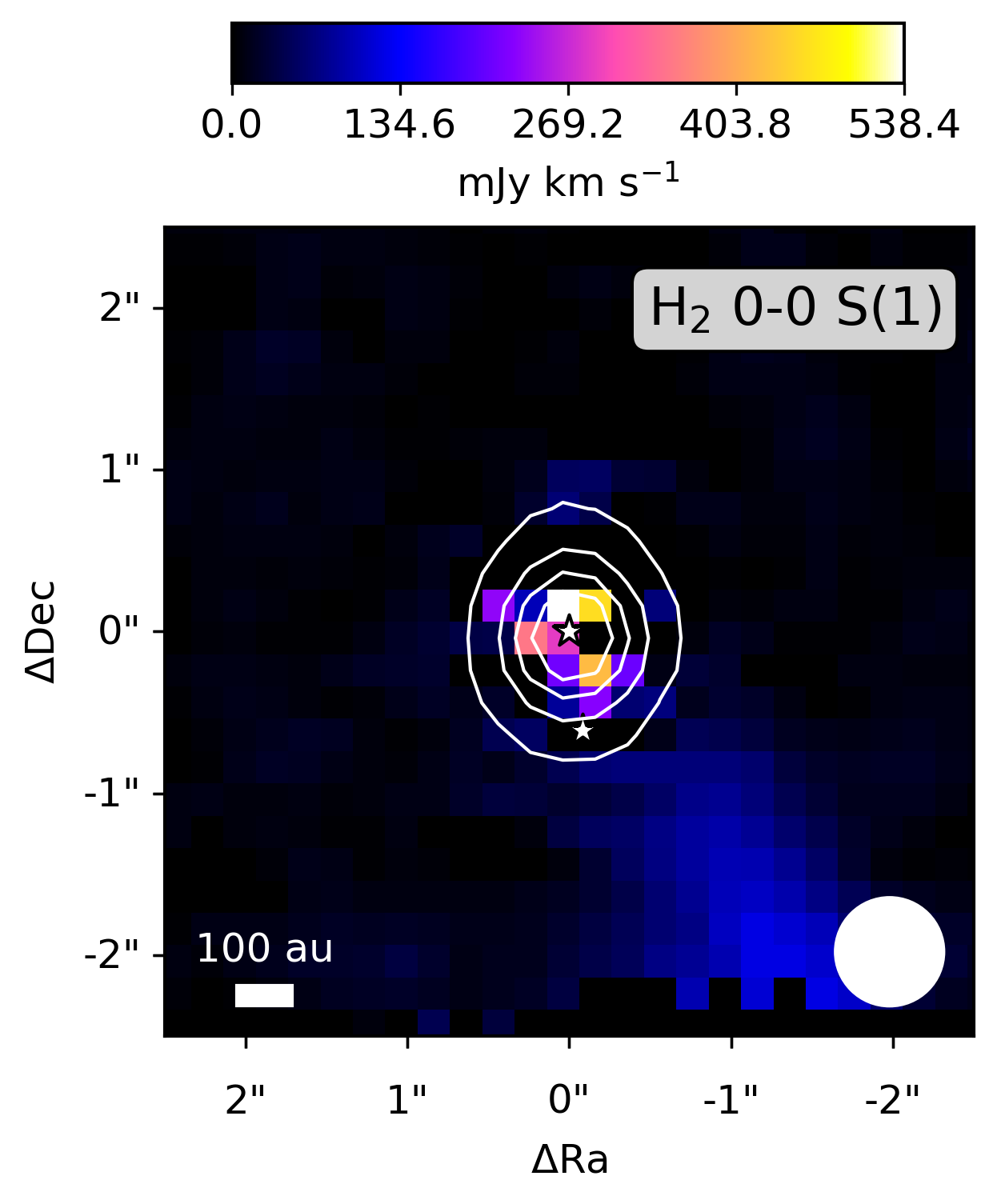}
    \includegraphics[width=0.45\linewidth]{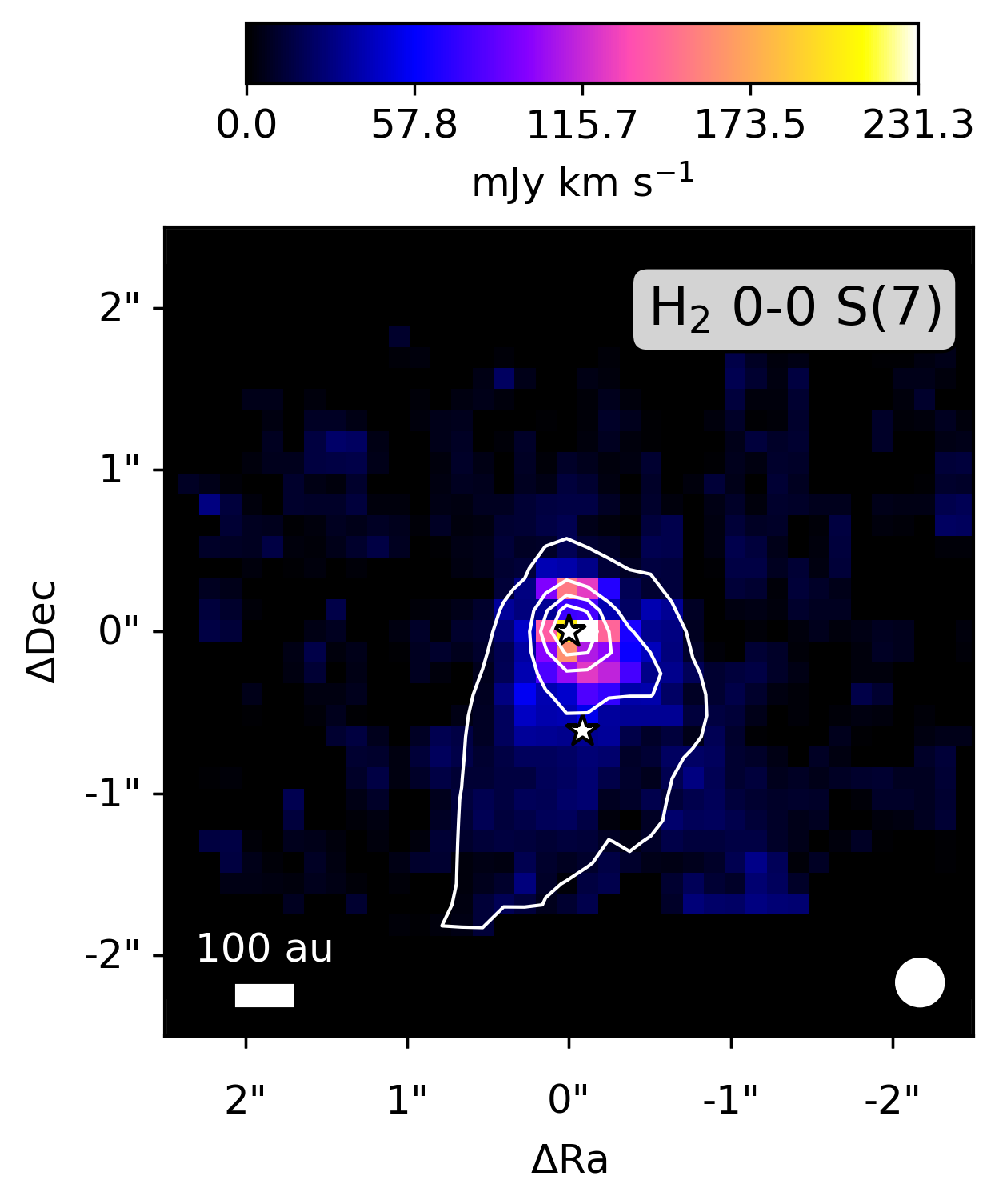}
    \caption{
    Integrated intensity map of the CO$_2$ $\nu_2=1-0$ $Q$-branch (top left), H$_2$O $\nu_2=1-0$ $4_{2,3}-5_{3,2}$ (top right), H$_2$ 0-0 S(1) (bottom left), and H$_2$ 0-0 S(7) (bottom right) observed with MIRI-MRS in color. The images are integrated over [14.9,15.0]~\mum for CO$_2$, [7.145-7.15]~\mum for H$_2$O, and [-0.01,0.01]~\mum with respect to the transitions of H$_2$. The extent of the continuum around the respective wavelengths is overlaid in white contours. A white scale bar is displayed in the bottom left of each panel and the size of the PSF is presented as the filled white circle in the bottom right. The direction of the two outflows originating from IRAS2A1 and IRAS2A2 are indicated with the colored arrows in the top left panel \citep{Tobin2015_IRAS2A}. The H$_2$O emission is peaking onsource on the same scales as SO$_2$, whereas the CO$_2$ and H$_2$ 0-0 S(1) emission are clearly peaking in the outflow toward the south-west. The H$_2$ 0-0 S(7) line peaks mostly onsource but also has a component in the south-western outflow.
    }
    \label{fig:co2_miri_map}
\end{figure}

\begin{figure*}[h]
    \centering
    \includegraphics[width=0.92\linewidth]{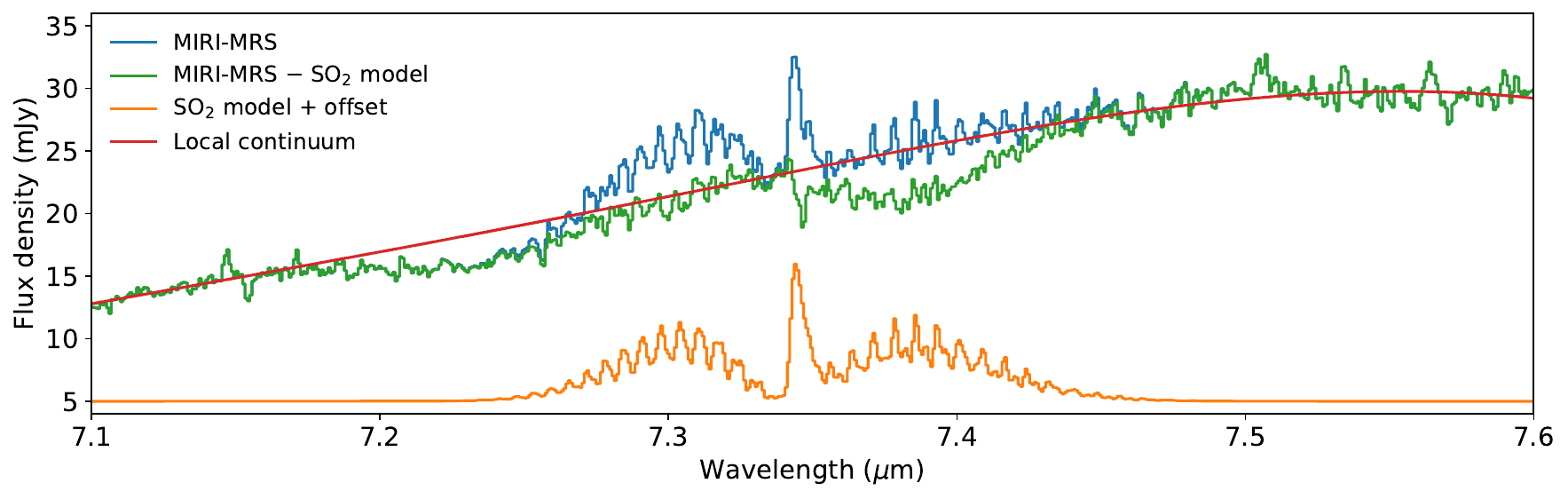}
    \caption{Spectrum of IRAS2A in blue centered around the $\nu_3$ band of SO$_2$ as observed by JWST/MIRI-MRS. The local continuum fit is shown in red and the best-fit LTE slab model is displayed in orange at a +5~mJy offset. In green, the SO$_2$ subtracted data is presented, revealing the 7.4~\mum ice absorption feature that was hidden by the $P$-branch of the SO$_2$ $\nu_3$ band. The $Q$-branch is slightly overfitted, resulting in a subtraction residual. The ice absorption bands are further analyzed by \citet{Rochasubm}.}
    \label{fig:so2_sub_spec}
\end{figure*}

\begin{figure*}[h]
    \centering
    \includegraphics[width=0.92\linewidth]{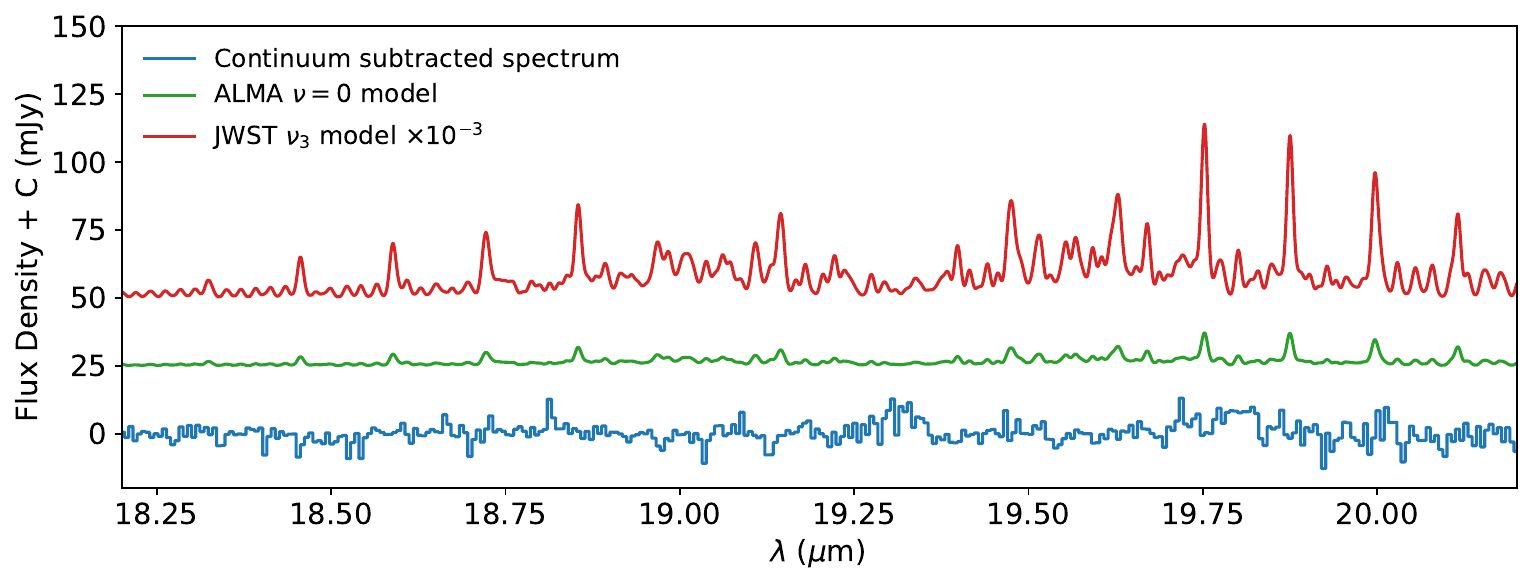}
    \caption{JWST/MIRI-MRS continuum subtracted spectrum (blue) centered on the $\nu_2$ band of SO$_2$ around 19~\mum. No clear emission or absorption features of SO$_2$ are detected. Overlaid are LTE slab models using the best-fit parameters derived from the $\nu_3$ band around 7.35~\mum (red) and the pure rotational lines in the ALMA data (green). The MIRI slab model is scaled down by a factor $10^{-3}$ for clarity. Both models are offset with respect to the data and are corrected for an extinction of $A_{\rm V} = 55$~mag \citep[][]{Rochasubm} using a modified version of the \citet{McClure2009} extinction law (see Appendix~\ref{app:ext_corr}).}
    \label{fig:so2_19mum}
\end{figure*}

\begin{figure*}[h]
    \centering
    \includegraphics[width=0.92\linewidth]{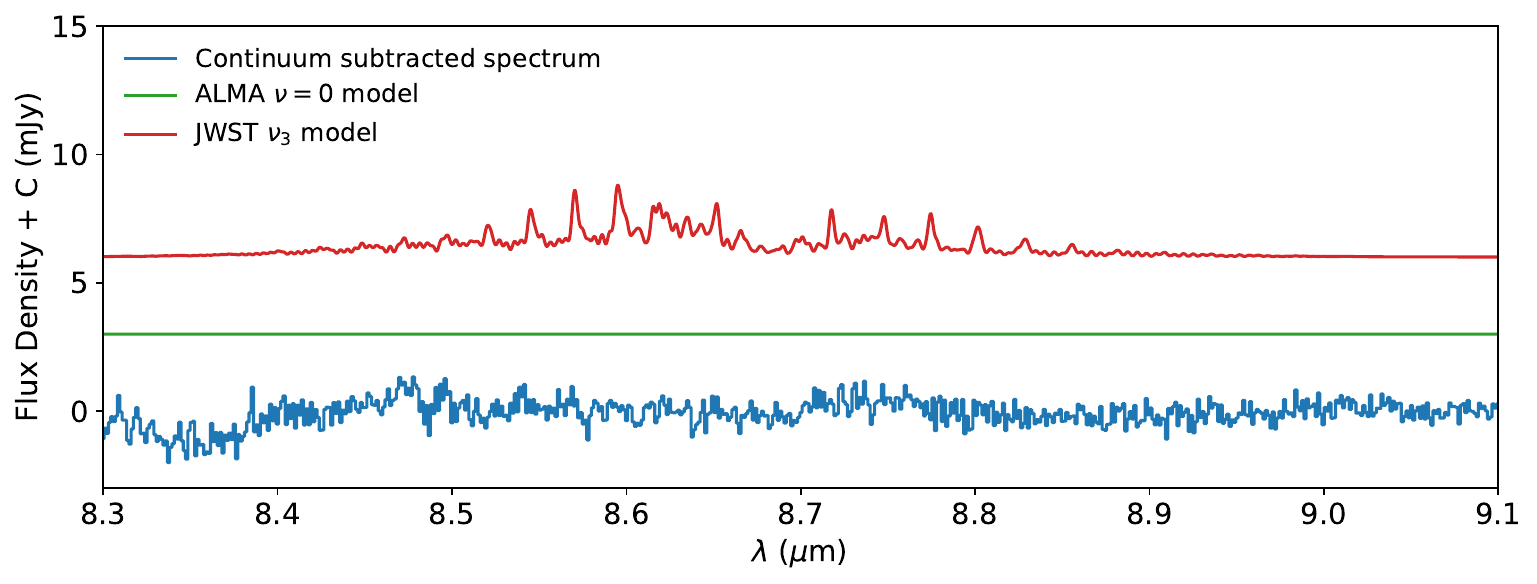}
    \caption{JWST/MIRI-MRS continuum subtracted spectrum (blue) centered on the $\nu_1$ band of SO$_2$ around 8.5~\mum. The two overlapping MIRI-MRS sub-bands (Channels 2A and 2B) have been stitched for clarity. No clear emission or absorption features of SO$_2$ are detected. Overlaid are slab models using the best-fit parameters derived from the $\nu_3$ band around 7.35~\mum (red) and the pure rotational lines in the ALMA data (green). Both models are offset with respect to the data and are corrected for an extinction of $A_{\rm V} = 55$~mag \citep{Rochasubm} using a modified version of the \citet{McClure2009} extinction law (see Appendix~\ref{app:ext_corr}).}
    \label{fig:so2_9mum}
\end{figure*}

\clearpage

\section{Extinction correction}
\label{app:ext_corr}
A modified version of the extinction law of \citet{McClure2009} is created for the extinction correction because the depths of the dominant absorption features (e.g., H$_2$O, CO$_2$, silicates) do not match with those measured toward IRAS2A. The total extinction (in units of optical depth) toward IRAS2A ($\tau_{\rm tot}(\lambda) = A_{\rm\lambda}$/1.086) is therefore decomposed into two components,
\begin{align}
    \tau_{\rm tot}(\lambda) = \tau_{\rm ice,silicate}(\lambda) + \tau_{\rm ext}(\lambda),
    \label{eq:tau_tot}
\end{align}
where $\tau_{\rm ice,silicate}(\lambda)$ is the differential extinction caused by the ice and silicate absorption features and $\tau_{\rm ext}(\lambda)$ is the absolute extinction. 

\begin{figure*}[t]
    \centering
    \includegraphics[width=\linewidth]{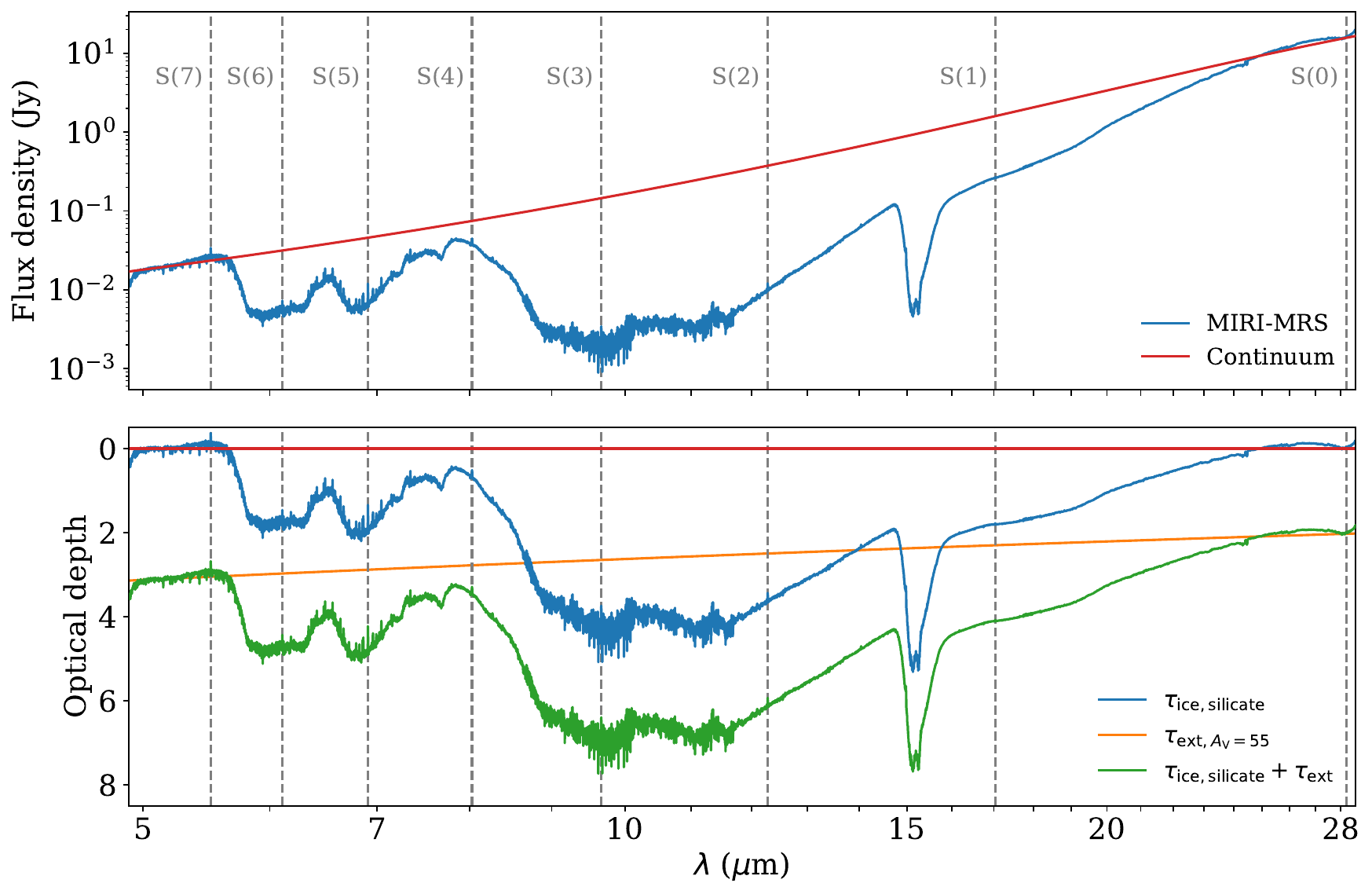}
    \caption{
    {\it Top}: observed MIRI-MRS spectrum of IRAS2A (blue) with the global continuum estimate (red) overplotted. The continuum is based on a third order polynomial fit to emission and absorption free wavelength ranges. The positions of H$_2$ $\nu=0$ rotational transitions are indicated with the gray vertical dashed lines. 
    {\it Bottom:} differential extinction in units of optical depth (blue) of the absorption features as function of wavelength derived from the global continuum estimate (Eq.~\eqref{eq:tau_ice}). Overplotted is the absolute extinction (orange) computed using Eq.~\eqref{eq:tau_ext}, see the main text for more details. The total extinction (green) is computed from the sum of the two (Eq.~\eqref{eq:tau_tot}). 
    } 
    \label{fig:cont_tau}
\end{figure*}

The differential extinction $\tau_{\rm ice,silicate}(\lambda)$ is obtained in a similar matter to what is used for typical ice analysis studies. The optical depth is computed with respect to a third-order polynomial fitted through obvious absorption free wavelengths \citep[i.e., 5.1, 5.3, 7.6, 22, 24~\mum;][]{Rochasubm}, see top panel of Fig~\ref{fig:cont_tau}. The optical depth can then be calculated via,
\begin{align}
    \tau_{\rm ice,silicate}(\lambda) = -\ln\left(\frac{\mathcal{F}_\lambda}{\mathcal{F}_{\rm cont}}\right),
    \label{eq:tau_ice}
\end{align}
where $\mathcal{F}_{\rm \lambda}$ is the measured flux with MIRI-MRS and $\mathcal{F}_{\rm cont}$ the continuum flux. The differential extinction in units of optical depth is presented in the bottom panel of Fig~\ref{fig:cont_tau}. 

\begin{figure*}[t]
    \centering
    \includegraphics[width=\linewidth]{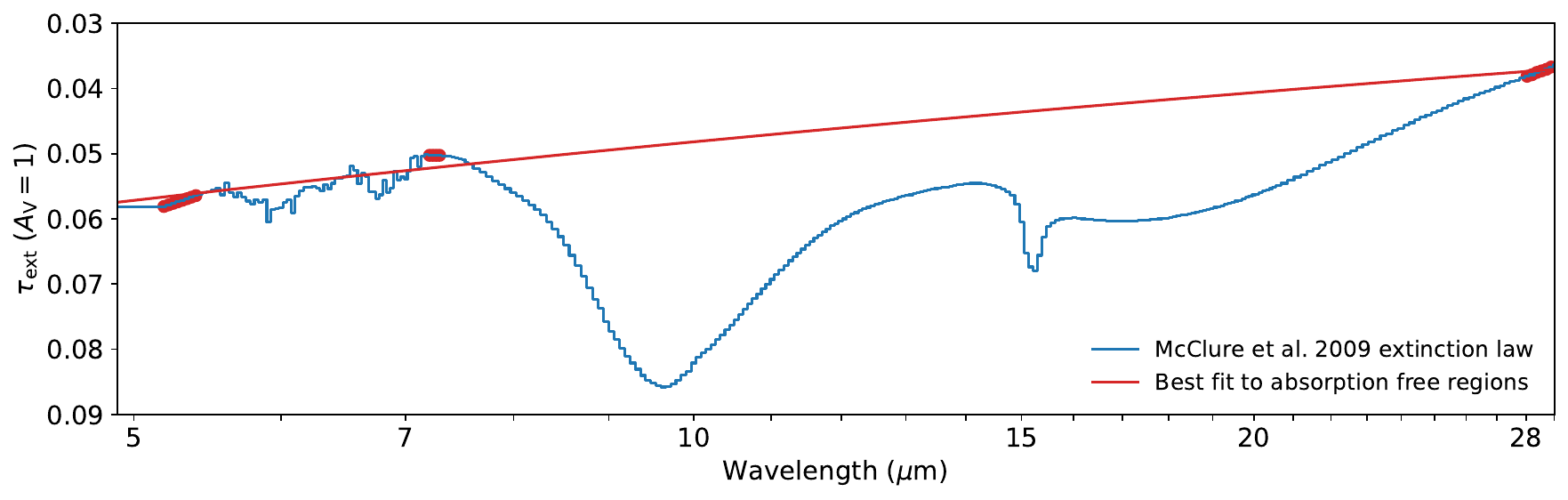}
    \caption{
   The extinction law of \citet[][]{McClure2009} for $A_{\rm V}=1$~mag (assuming $A_{\rm K} = A_{\rm V}/7.75$~mag) in blue with the best-fit powerlaw model to the absorption free regions (dots) overplotted in red. Both the extinction law and the best-fit model are plotted in units of optical depth for $A_{\rm V}=1$~mag, but are directly proportional to $A_{\rm V}$.
    } 
    \label{fig:mcclure_extinction}
\end{figure*}

The absolute extinction $\tau_{\rm ext}$ is computed by fitting a powerlaw model to the extinction law of \citet{McClure2009} for an extinction of $A_{\rm K} > 1$. Only the wavelengths outside of the major absorption features are taken into account (i.e., 4.9-5.4,7.2-7.3,28-29~\mum). In this case, the power of the powerlaw is also fitted. The extinction law is fitted for the case of $A_{\rm V}=1$~mag \citep[assuming $A_{\rm K}=A_{\rm V}/7.75$~mag;][]{McClure2009} in units of optical depth ($\tau_{\rm ext}(\lambda) = A_{\rm\lambda}$/1.086). The fit to the extinction law is presented in Fig.~\ref{fig:mcclure_extinction}. 
Using this fit, the absolute extinction is derived to be, 
\begin{align}
    \tau_{\rm ext}(\lambda) = 0.085 \lambda^{-0.25} A_{\rm V}.
    \label{eq:tau_ext}
\end{align}
Here, $A_{\rm V} = 55$ is adopted based in the depth of the silicate absorption \citep[][]{Rochasubm}, leading to $\tau_{\rm ext}\sim3$ around 5~\mum and $\tau_{\rm ext}\sim2$ around 25~\mum (see bottom panel of Fig~\ref{fig:cont_tau}). The total extinction toward IRAS2A in units of optical depth is presented in the bottom panel of Fig.~\ref{fig:cont_tau}.

\section{H$_2$ analysis}
\label{app:H2_analysis}
\subsection{Fitting of H$_2$ lines}
\begin{figure*}[h]
    \centering
    \includegraphics[width=\linewidth]{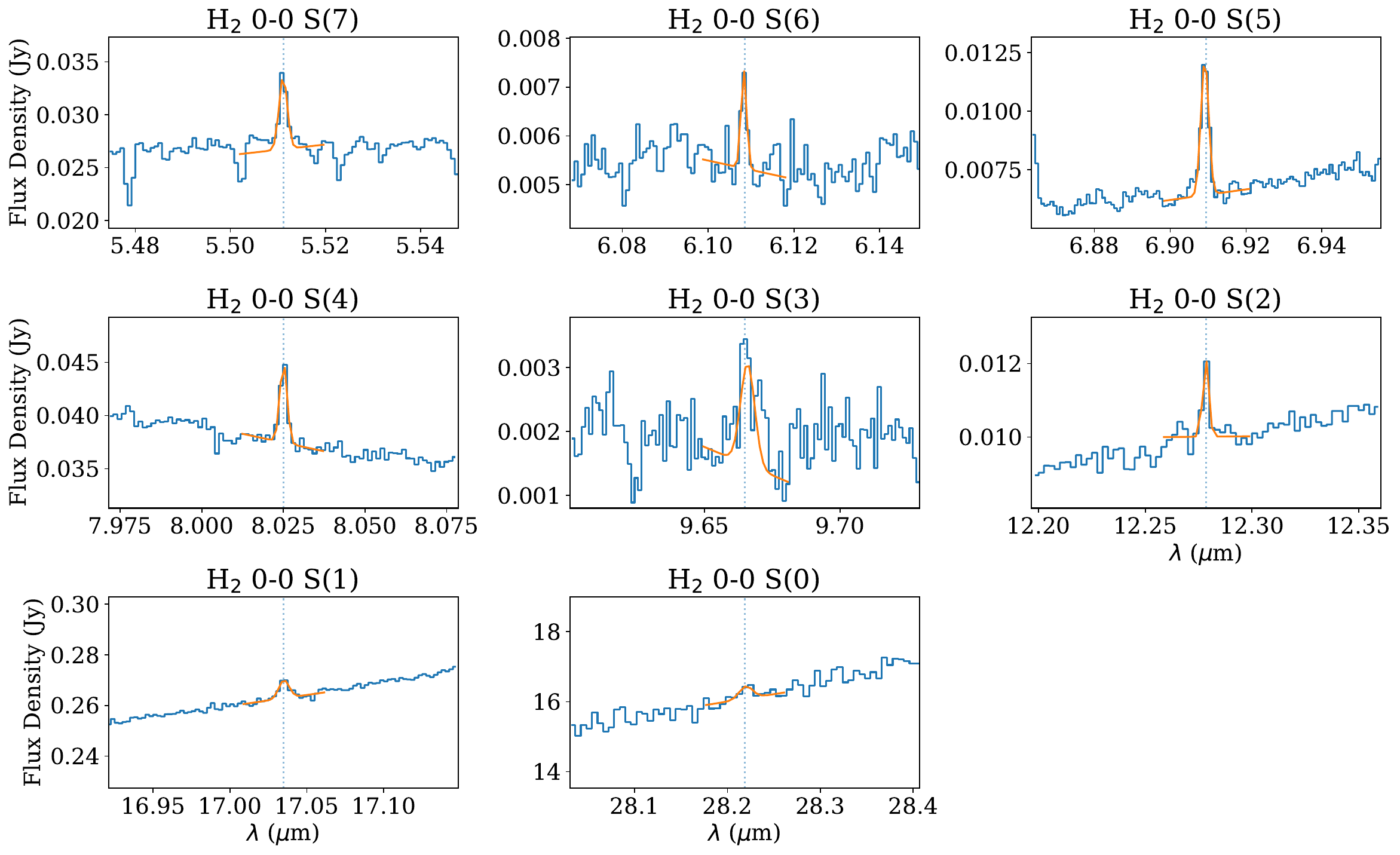}
    \caption{Observed MIRI-MRS spectrum (blue) centered around H$_2$ $\nu=0$ pure rotational lines from the S(7) line in the top left panel to the S(0) line in the bottom middle panel. The best-fit Gaussian model is presented in orange in each panel. All presented lines are considered detected except for the S(0) line since the flux is similar to that of several strong residual fringes located in the vicinity of the S(0) line.} 
    \label{fig:H2_line_fits}
\end{figure*}

The MIRI range covers the $\nu=0$ pure rotational lines of H$_2$ from the S(0) line at 28.22~\mum till the S(8) line at 5.05~\mum. However, the S(0) line lies at the very edge of Channel 4C ($\lambda>24$~\mum) where the sensitivity drops significantly and the flux calibration is very poor. Some weak line emission appears to be present (see Fig.~\ref{fig:H2_line_fits} bottom middle panel) but since the peak intensity is equally strong as that of residual fringes present surrounding the line, it is excluded form the rotational diagram analysis. All other H$_2$ $\nu=0$ transitions are detected except for the S(8) transition. The detected transitions are fitted with a simple Gaussian emission profile (see Fig.~\ref{fig:H2_line_fits}). A flux calibration uncertainty of 5\% is assumed. The H$_2$ line fluxes are corrected for total extinction based on a modified version of the extinction law of \citet[][]{McClure2009} introduced in Appendix~\ref{app:ext_corr}
The integrated line fluxes are reported in Table~\ref{tab:H2_fluxes}.

\renewcommand{\arraystretch}{1.2}
\begin{table}[h]
    \centering
    \caption{Integrated line fluxes of H$_2$ transitions.}
    \label{tab:H2_fluxes}
    \begin{tabular}{cccccccc}
    \hline\hline
    Transition & $\lambda$ & $E_{\rm up}$ & $g_{\rm u}$ & $\mathcal{F}_{\rm line}$ & $\tau_{\rm tot}$\tablefootmark{1} & $\mathcal{F}_{\rm line}\times\exp\left(\tau_{\rm tot}\right)$ & $N_{\rm u}$ \\
    & \mum & K & & W m$^{-2}$ arcsec$^{-2}$ & & W m$^{-2}$ arcsec$^{-2}$ & cm$^{-2}$ \\
    \hline
    0-0 S(7) & \,\,\,5.511 & 7196.7 & 57 & 4.5$\pm$1.4(-17) & 2.68 & 4.3$\pm$1.3(-16) & 3.2$\pm$1.6(18) \\
    0-0 S(6) & \,\,\,6.109 & 5829.8 & 17 & 9.9$\pm$3.4(-18) & 4.44 & 5.5$\pm$1.9(-16) & 7.8$\pm$4.3(18) \\
    0-0 S(5) & \,\,\,6.910 & 4586.1 & 45 & 4.5$\pm$0.4(-17) & 4.25 & 2.1$\pm$0.2(-15) & 6.5$\pm$1.9(19) \\
    0-0 S(4) & \,\,\,8.025 & 3474.5 & 13 & 6.3$\pm$0.5(-17) & 3.29 & 1.1$\pm$0.2(-15) & 9.1$\pm$2.6(19) \\
    0-0 S(3) & \,\,\,9.665 & 2503.7 & 33 & 3.1$\pm$1.3(-17) & 6.39 & 1.2$\pm$0.5(-14) & 3.2$\pm$2.0(21) \\
    0-0 S(2) & 12.279 & 1681.6 & \,\,\,9 & 2.3$\pm$0.6(-17) & 5.94 & 5.6$\pm$1.4(-15) & 6.7$\pm$3.1(21) \\
    0-0 S(1) & 17.035 & 1015.1 & 21 & 1.8$\pm$0.4(-16) & 4.08 & 7.0$\pm$1.4(-15) & 6.8$\pm$2.7(22) \\
    \hline
    \end{tabular}
    \tablefoot{$a(b)$ means $a\times10^b$. 
    \tablefoottext{1}{Total extinction in units of optical depth at the respective wavelength $\lambda$ computed using Eq.~\eqref{eq:tau_tot}. The line optical depths of the H$_2$ transitions themselves are all small ($\tau_{\rm line}\ll1)$. }
    }
\end{table}
\renewcommand{\arraystretch}{1.0}

\subsection{Fitting the rotational diagram}
\begin{figure}[h]
    \centering
    \includegraphics[width=0.6\linewidth]{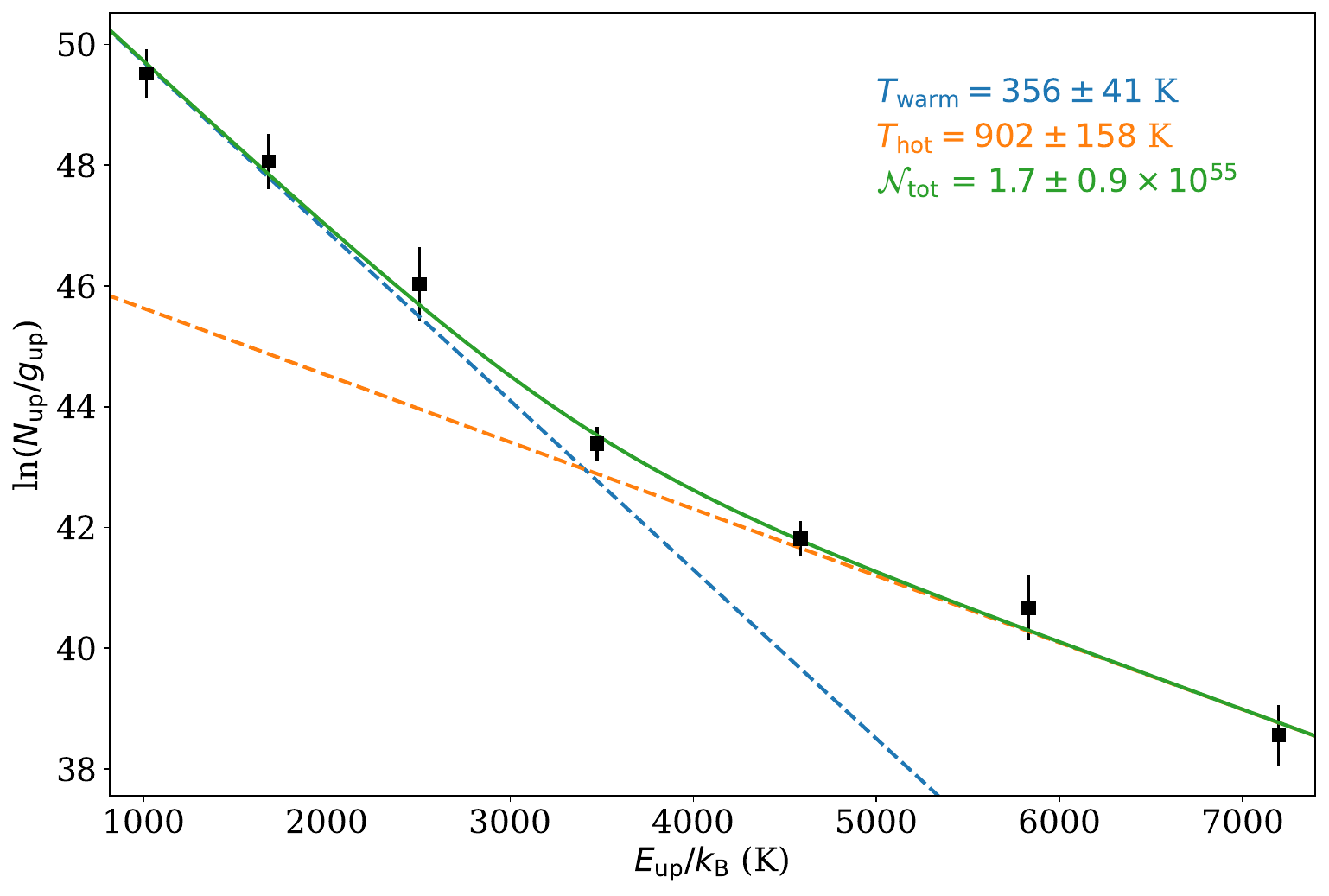}
    \caption{Rotational diagram of H$_2$ derived from the MIRI-MRS data. The black datapoints indicate the measured fluxes which are corrected for extinction (see Table.~\ref{tab:H2_fluxes}). The best-fit two-component model is presented as the solid green line and the two individual components are shown as the blue (warm) and orange (hot) dashed lines. The best-fit parameters are displayed in the top right.} 
    \label{fig:H2_rot_diag}
\end{figure}

A rotational diagram is created following the formalism of \citet[][]{Goldsmith1999} and presented in Fig.~\ref{fig:H2_rot_diag}. 
The rotational diagram can be best fitted using a two-component model: a warm component with a rotational temperature of $T_\mathrm{warm}=356\pm41$~K and a column density of $N_{\rm warm} = 5.6 \pm 2.9 \times 10^{23}$~cm$^{-2}$ and a hot component with $T_\mathrm{hot} = 902 \pm 158$~K and a column density of $N_{\rm hot} = 4.4 \pm 5.1 \times10^{21}$~cm$^{-2}$. The total number of molecules can be constrained to $\mathcal{N_{\rm H_2}} = 1.7 \pm 0.9 \times 10^{55}$ molecules. An emitting area with a diameter of $1.4''$ is used (i.e., $R_{\rm source} = 205$~au at a distance of $293$~pc), equal to the size of the aperture at 7.35~\mum. However, since the H$_2$ emission is assumed to be optically thin, the value of $\mathcal{N}_{\rm tot}$ does not depend on the assumed emitting area.

\end{document}